\documentclass[10pt, aps, pra, 
twocolumn, notitlepage,superscriptaddress,nofootinbib,floatfix   
]{revtex4-2}

\usepackage[T1]{fontenc}
\usepackage[utf8]{inputenc}
\usepackage{textcomp}  
\usepackage{amsmath,amssymb,mathtools}
\usepackage{lmodern}
\usepackage{graphicx}
\usepackage{upgreek}
\usepackage{parskip}
\usepackage[version=4]{mhchem} 
\usepackage{xcolor} 

\usepackage{hyperref}
\hypersetup{%
	pdfpagemode=FullScreen,
	pdfstartpage=1,
	pdfmenubar=true,
	pdftoolbar=true,
	colorlinks = true,
	linkcolor=blue,
	citecolor=blue,
	urlcolor=blue,
	bookmarksopen=false
}

\begin{document}

\title{Scattering of a massive quantum vortex-dipole from an obstacle}

\author{Alice Bellettini} 
\email{alice.bellettini@polito.it}
\affiliation{Department of Applied Science and Technology, Politecnico di Torino, 10129 Torino, Italy}

\author{Enrico Ortu} 
\affiliation{Department of Mathematical Sciences
"Giuseppe Luigi Lagrange", Politecnico di Torino, 10129 Torino, Italy}
\affiliation{CONCEPT Lab, Fondazione Istituto Italiano di Tecnologia, Via E. Melen
83, Genova, 16152, Italy}
\affiliation{INFN, Sezione di Torino, Via P. Giuria 1, 10125 Torino, Italy}

\author{Vittorio Penna}
\affiliation{Department of Applied Science and Technology, Politecnico di Torino, 10129 Torino, Italy}

\date{\today}

\begin{abstract}
   In binary mixtures of Bose-Einstein condensates, massive-vortex dipoles can arise, and undergo scattering processes against obstacles. These show an intriguing dynamics, governed by the strongly nonlinear character of the quantum vortex motion, where we are able to highlight the effects of the boundaries. We first characterize such scattering dynamics via some point-like models, for the cases of an unbounded plane and a confined geometry. Within
   this framework, we find two fundamental scattering behaviors of a vortex dipole, the ``fly-by'' and the ``go-around'' processes.
   By plotting the deflection angle of the dipole versus the impact parameter we are able to quantify the transition between different scattering behaviors.
   We then are able to introduce an analytical distinction of the two scenarios, basing on the point-like model for the plane geometry.
   Furthermore, another interesting result shows the emergence of an on-average massless dynamics
   whenever the nonlinear interactions with the obstacle become negligible.
   Alongside, we investigate the quantum dipole scattering via the numerical simulation of two coupled Gross-Pitaevskii equations, describing the quantum mixture at a mean-field level. 
   In this way, we benchmark the point-like model against the mean-field simulations. 
\end{abstract}

\maketitle

\section{Introduction}

The idea that two-dimensional quantum vortex \cite{Onsager1949, Feynman1955} dipoles, in their free-particle-like motion \cite{Penna1999}, can undergo scattering phenomena against
obstacles in the system is not only of experimental relevance but also intriguing in that the phenomenon lays at the basics of two-dimensional quantum turbulence. The understating and modeling of two-vortex systems opens the path to a broader phenomenology in the direction of the effects of an obstacle in a many-dipoles systems. Also, an accurate description of the trajectories of scattering vortex-dipoles, being they relatively robust configurations, allows for the indirect control of these structures' trajectory.
Finally, a third aspect that motivated this work is the effect of a vortex mass, combined with the choice of initial conditions, on the (scattering) dynamics of vortex dipoles. 

Massive vortices \cite{Law2011, Bandyopadhyay2017, Richaud2020} arise in the context of mixtures of Bose condensed gases \cite{KASAMATSU2005}, and are part of a rich phenomenology distinguishing multi-component systems with respect to single-component ones, thanks to the interplay of different type of interactions and their tunability. 
We refer to it as a massive vortex when a quantum vortex, hosted by a majority component ``$a$'' exhibits a filled, enlarged, core due to a second minority component ``$b$'', while the $b$-density outside of the vortex core is negligible. These structures, whose stability properties were studied in Refs. \cite{McGee2001, Garcia2000, Skryabin2000},
are made robust by the immiscibility condition \cite{Colson1978, Trippenbach2000} of the two components. The dynamics of massive vortices, endowed with effective inertial effects \cite{Richaud2020}, is characterized by a broader class of solutions with respect to massless vortices. In fact, massive vortices are associated in point-like approximations to second-order equations of motion \cite{Richaud2021}, whereas massless vortices
are governed by first-order equations \cite{Helmholtz1858, Kirchhoff1876, Saffman1993}. The effect of the infilling mass on well separated vortices is not only to modify the phase space of the system, but also to introduce new types of vortex trajectories, characterized by some oscillations
around an average trajectory \cite{Bellettini2024}.
On the other hand, close enough vortices make up effective double-well potentials for their infilling component. In this case, the quantum tunneling of the infilling component can take place, coupling the bosonic transfer to the dynamics of the quantum vortices \cite{Bellettini2024PRR}. Novel effects such as vortex-supported supercurrents were found in these systems.
Most recently, the concept of massive vortex became relevant also within the physics of single-component Fermi gases,
where
the appearance of a vortex mass was investigated
\cite{Levrouw2025}.

The first observations of quantum vortices in a Bose-Einstein condensate (BEC) were due to the experiments of Refs. \cite{Matthews1999, Anderson2000, Madison2000, Raman2001, Abo-Shaeer2001} in 1999 and the early 00s.
In particular, in the experiments by Anderson \textit{et al.} \cite{Anderson2000} and by Matthews \textit{et al.} \cite{Matthews1999} quantum vortices with filled and empty cores
were observed in a mixture of two hyperfine components of $^{87}\mathrm{Rb}$.
Subsequently, vortex dipoles in Bose-Einstein condensates were experimentally realized by Neely \textit{et al.}
\cite{Neely2010} 
and by Freilich \textit{et al.} \cite{Freilich2010}, where their real-time dynamics was also observed. 
Later, Kwon \textit{et al.} \cite{Kwon2021} also created quantum vortex dipoles, in a fermionic gas,  
and studied their collision and relation to sound waves emission.

In the theoretical literature, the scattering of massless vortex dipoles against an obstacle was analyzed by Penna \cite{Penna1999} by applying the dynamical-algebra method \cite{SOL1971,PR2017}
to a vortex pair with arbitrary vortex charges.
Later, Griffin \textit{et al.} \cite{Griffin2017} numerically studied the dynamics of massless vortex dipoles when scattered against an impurity or third vortex. While their results can be directly compared to our work and are in agreement with it, we extend our study
to massive vortices and 
by providing an analytical framework and an approximation scheme that allows for the discrimination of the different scattering behaviors.

In this paper we investigate the scattering process of a dipole of massive quantum vortices caused by the presence of a disk-like obstacle. To capture the essential properties characterizing the dipole scattering, we start with considering a system in a large planar domain and postpone to a subsequent part the investigation of the more realistic case where a circular boundary, representing an external potential trapping the system, is included. The study of the scattering is based on a well-known point-like model (PLM) derived within a variational approach from the field Lagrangian of a binary mixture \cite{Kim2004,Richaud2021}.
Importantly, the scattering scenario, involving the disruption and recombination of the vortex-dipoles, challenges the validity of such point-like models more than what done in previous literature \cite{Bellettini2024}. 
Among many interesting aspects, the PLM reveals how the dipole dynamics essentially features two possible behaviors: either the dipole is deflected by the obstacle (fly-by case) or the dipole undergoes a splitting effect where a vortex
is captured by the obstacle and, after circumnavigating it, recombines in a dipole configuration (go-around case). 
Interestingly, the robustness of the dipole structure evidenced by our numerical simulations can be validated at the analytical level through a perturbative treatment of the vortex-pair PLM equations. The latter, in fact, disclose the existence of massless-like asymptotic regimes (far enough from the obstacle) featuring a dipole-like behavior. The complex interplay among the intervortex distance, the scattering impact parameter and the vortex masses that can lead to the dipole splitting
is then analyzed both via numerical simulations of the PLM equations, and through an analytical approach allowing us to find the critical condition that separates the fly-by behavior from the go-around behavior. 
Moreover, we show that it is possible to extract the scattering deflection angle of a dipole as a function of the scattering impact parameter, thus implying an indirect control of the dipole's trajectory via its position. 

Thanks to the inclusion a circular boundary, we then move on to the description of the scattering dynamics in an annulus geometry. We investigate the effect of confinement on the vortex-antivortex (VA) dynamics showing how the scattering process due to the central obstacle is quantitatively modified by the presence of the boundary, while its qualitative features persist.
The
dipole behavior remains a robust feature of the VA pair far enough from the central obstacle and the confining boundary. 
The latter introduces a new distinctive feature of the bounded system, that is the possibility to trigger a different type of dipole splitting at the circular boundary. In this case, the vortex and the antivortex are captured by their virtual counterparts (describing the effect of the boundary) thereby creating two virtual dipoles that travel along the boundary. Our analysis aims at offering an insight into the complex dynamics of vortex dipoles, which are found to exhibit almost periodic
pair separations, and recombinations followed by the scattering from the central obstacle.
In spite of the more complex geometry of an obstacle within a disk, 
this rich dynamical scenario can be validated
by the numerical simulation of the mean-field Gross-Pitaevskii equations, representing the field picture of the mixture dynamics \cite{Gross1961, Pitaevskii1961}. Remarkably, it turns out that the solutions of the PLM are able to well capture this mean-field dynamics also within this more complex picture.

This paper is structured as follows. In Section \ref{sec:PL_model} we introduce the point-like models of interest for the modeling of scattering processes in a large planar domain and in a two-dimensional (2D) disk. We then proceed in Section \ref{sec:PLM_plane} with illustrating and analyzing the scattering behaviors of a vortex dipole, as predicted by the point-like model for the very large domain. 
In Section \ref{sec:scatt} we derive analytically the condition distinguishing the two scattering behaviors. Subsequently, in Section \ref{sec:PLM_annulus}, we move on to the description of the scattering dynamics in the disk-like geometry. This more realistic case is then validated against numerical simulations of the mean-field Gross-Pitaevskii equations in Section \ref{sec:GPE}. 
After discussing the effectiveness of the PLM dynamics, we consider the beyond-point-like effects. 
Finally, in Section \ref{sec:concl}, we provide a final summary and discuss the most interesting future outlooks.

\section{Point-like models}
\label{sec:PL_model}

A widely investigated setting sees $N_v$ vortices within a very large domain which can be confused with an infinite plane. In this case, the effects of a confining potential are neglected, or equivalently the vortices are close enough to the center of the trap as compared to their distances from the boundary. 
In such system, it is then interesting to introduce a disk-like obstacle centered in the origin, and study the scattering process induced by it on the dynamics of a vortex dipole. We choose to model such an obstacle by an external potential, realizable in the atomic mixture by means of auxiliary laser beams (see e.g. the experiment in Ref. \cite{Samson2016}). In this case, the border of the obstacle can be considered as an effective internal boundary, induced by a step-like potential.

The dynamics of $N_v$ planar massive vortices
can be described via point-like models.
These are obtained
via a time-dependent variational approach starting from the following mean-field Lagrangian, involving the $a$- and $b$-condensate wavefunctions $\psi_a\coloneq \psi_a(\vec{r}, t)$ and $\psi_b\coloneq \psi_b(\vec{r}, t)$,

\begin{equation}
\begin{split}
	\mathcal{L}
    &=\sum_{i=a,\;b} \bigg[ \frac{i\hbar}{2}\int \mathrm{d}^2 r\bigg(\psi_i^*\dot\psi_i  - \dot \psi_i^* \psi_i\bigg) 
    \\
	&-\int 
\int\mathrm{d}^2r\bigg(\frac{\hbar^2}{2m_i}|  \nabla \psi_i|^2 
+ V_{ \text{ext}}(\vec{r})|\psi_i|^2 
\\
&+ \frac{g_i}
{2d_z}|\psi_i|^4\bigg) \bigg ]
     - \frac{g_{ab}}{d_z} \int\mathrm{d}^2r\; |\psi_a|^2|\psi_b|^2,
\end{split}
\label{eq:L_GP}
\end{equation}

\medskip
\noindent
with 
$g_{ab}=2\pi \hbar^2 a_{ab}/m_r$, where $a_{ab}$ denotes the $s$-wave scattering length relevant to the inter-component interactions, and the reduced mass $m_r$ is such that $1/m_r=1/m_a+1/m_b$, with $m_a$ ($m_b$) the atomic mass of component $a$ ($b$).
The intra-component interaction coupling constants are
$g_a=4\pi \hbar^2 a_a/m_a$ and $g_b=4\pi \hbar^2 a_b/m_b$, depending respectively on the $s$-wave scattering lengths $a_a$ and $a_b$, relevant to the component $a$ and the component $b$. The interaction parameters are normalized by the quantity $d_z$, that is a virtual thickness in the third dimension of the 2D system, and is needed for the dimensionality reduction of the Lagrangian \cite{Salasnich2002}.
Finally, in Eq. \eqref{eq:L_GP}, $V_{ \text{ext}}(\vec{r}) $ indicates the external potential.
This consists of a small disk-like hard-wall potential, located at the center of the trap, and mimicking the presence of an obstacle. Note that, due to the rotational symmetry featured by our systems, the angular momentum is conserved.

The Lagrangian system conserves the number of particles in the two components, given by 

$$N_j=\int d^2 r\, |\psi_j|^2,\;\;\; j=a,b.$$

The following \textit{Ans{\"a}tze} for the condensates wavefunctions are plugged into the Lagrangian \eqref{eq:L_GP} to derive the corresponding point-like model,
\begin{equation}
    \label{eq:psi_a}
    \psi_a=\sqrt{n_a 
    }\,e^{i\sum_{j} N_j \left[\arctan\left(\frac{y-y_j}{x-x_j}\right)-\arctan\left(\frac{y-y_j^\prime}{x-x_j^\prime}\right)\right]},
\end{equation}
and
\begin{equation}
\label{eq:psi_b}
\psi_b  =  \left(\frac{N_b}{N_v\pi\sigma_b^2}\right)^{1/2}\sum_je^{-|\vec r - \vec r_j|^2/2\sigma_b^2}\,e^{i\vec r\cdot \vec \alpha_j},
\end{equation}

\medskip
\noindent
where the \textit{Ans{\"a}tze} \eqref{eq:psi_a} and \eqref{eq:psi_b}
hold in the regions where $V_{\text{ext}}(\vec r)=0$ and are zero elsewhere. In the above, $N_j=\pm 1$ is
the $j$-th vortex charge, while
$n_a$ describes the average $a$-particle density, and takes a constant value almost everywhere except at
the $a$-phase singularity points, in correspondence of the vortex centers, where $n_a$ is taken to be zero. In this way, the effects of the vortex-core shapes are neglected, approximation which holds when the pair-distance between any two vortices is large enough compared to the core's size. As mentioned, the \textit{Ansatz} for $\psi_a$ contains phase-singularity points at the quantum vortex centers, expressed by the Cartesian coordinates vectors $\vec{r}_j=(x_j,y_j)^T$ and $\vec{r}_j\,'=(x'_j,y'_j)^T$, $j=1,...,N_v$. These are respectively the set of the real vortices in the system, and the set of the corresponding virtual image-vortices that, within the virtual-charge method \cite{VCM}, allow for a description of the effects of the hard-wall external potential $V_{\text{ext}}(\vec r)$.
In the current geometry, where the obstacle is represented by a circular boundary, this method entails $\vec{r}\,'_j =R_1^2 \vec{r}_j/r_j^2$. On the other hand, the \textit{Ansatz} for $\psi_b$ describes a set of Gaussian density-peaks (of width $\sigma_b$ and local velocity $ \hbar/m_b\, \vec{\alpha}_j$) representing the vortex infilling masses. The parameters $N_a$ and $N_b$ indicate the total number of $a$ and $b$-atoms, respectively, in the system.

The PLM obtained via the \textit{Ans{\"a}tze} \eqref{eq:psi_a} and \eqref{eq:psi_b},  
including a
disk-like obstacle of radius $R_1$ via the virtual-charge method,
is characterized by the
following Lagrangian \cite{Kim2004, Richaud2021} $L$, expressed in terms of the polar coordinates $(r_j,\theta_j)$ of the $j$-th vortex, $j=1,...,N_v$,

$$
    L = \sum_j 
 \frac{k_j m_a n_a}{2}\left(  R_1^2  -r_j ^2\right) \dot{\theta}_j +   \sum_j \frac{m_j}{2}\left( r_j ^2 \dot{\theta}_j ^2 + \dot{r}_j ^2 \right)\\
$$
\begin{equation}
- \sum_j W_j  - \sum_{j}\sum_{k\neq j} V_{jk},
\label{eq:Lagr}
\end{equation}

\medskip
\noindent
with ${V}_{jk} ={\bar U}_{jk}+ {U}_{jk}$,

\begin{equation}
 {\bar U}_{jk} = \frac{k_j k_k \, \rho_*}{8\pi} \ln\left( 
\frac{R_1^2 - 2   r_j  r_k  \cos \theta_{jk} + r_j ^2 r_k ^2/R_1^2 }
{ R_1^2 }
\right) ,
\label{eq:V_plane}
\end{equation}

\begin{equation}
U_{jk} = \frac{k_j k_k \, \rho_*}{8\pi}
\; \ln\left( 
\frac{ R_1^2}
{ r_j ^2 - 2 r_j  r_k \cos\theta_{jk}  + r_k ^2  } 
\right ),
\label{eq:RV_plane}
\end{equation}
where $\rho_* = m_a n_a$ is the mass density, 
${\bar U}_{jk}$ represents the interactions between the set of real vortices and antivortices with the set of their virtual counterparts, and ${U}_{jk}$ 
represents the interactions among real vortices or antivortices. Note that $j \ne k$ as 
the interaction of each vortex with its virtual counterpart is described by

\begin{equation}
   W_j = \frac{k_j^2 \, \rho_*}{4\pi} \ln\left(1 - \frac{r_j^2}{R_1^2} \right) .
   \label{eq:phi_plane}
\end{equation}

In the above, we introduced the notation $\theta_{jk} \coloneq\theta_j  - \theta_k$, while $m_j$
denotes the single core's mass.
Also, $k_j=N_jh/m_a$ is the vorticity of the $j$-th vortex, with $h$ the Planck's constant.

Within the PLM description, the dynamical equations are derived by means of Euler-Lagrange equations and read
\begin{equation}
m_j \ddot{r}_j  =  m_j r_j \dot{\theta}_j ^2 
- k_j \rho_* r_j
\dot{\theta}_j
- \frac{\partial W_j}{\partial r_j}-\sum_k \frac{\partial V_{jk}}{\partial r_j}  ,
\label{eq:eq_rho}
\end{equation}

\begin{equation}
m_j r_j ^2 \ddot{\theta}_j  =
k_j \rho_* r_j \dot{r}_j  - 
2 m_j r_j \dot{\theta}_j \dot{r}_j - \sum_k\frac{\partial V_{jk}}{\partial \theta_j} .
\label{eq:eq_theta}    
\end{equation}

While the geometry of a circular obstacle in the infinite plane, incorporated in Equations \eqref{eq:eq_rho} and
\eqref{eq:eq_theta}, will allow us to enact an
analysis focused on the scattering processes of a vortex dipole, the inclusion of an external confining boundary provides in fact a more realistic model. This 
takes into account the effect of a trapping potential, and allows for the comparison of the PLM results with the numerical results obtained from the Gross-Pitaevskii equations (GPEs, see Section \ref{sec:GPE}).
This geometry is described by the annulus model, which sees the 
vortices confined in a disk of radius $R_2$ and scattered against the central circular obstacle of size $R_1$. In this case, $V_{ \text{ext}}(\vec{r}) $ in Eq. \eqref{eq:L_GP} includes not only the obstacle contribution, but also the additional trapping hard-wall circular potential.
As the introduction of a boundary adds up complexity to the system, we can reasonably expect that the larger the domain is, the closer is the agreement between GPEs and PLM. Note that, in the limit $R_2\to+\infty$ the PLM for the disk with the central obstacle tends to the PLM in the infinite plane,
associated to Eqs. \eqref{eq:V_plane}-\eqref{eq:phi_plane}.

The PLM Lagrangian describing $N_v$ vortices within an effective annular domain is formally the same as the Lagrangian \eqref{eq:Lagr},
and the equations of motion are formally
the same as Eqs. \eqref{eq:eq_rho} and \eqref{eq:eq_theta}.
However, in this case
the single-vortex boundary-interaction 
term $W_j$ 
and vortex pair-interaction term $V_{jk}$ \cite{Caldara2023}
are

\begin{equation}
W_j
=\frac{k_j^2 \rho_*}{4\pi}
\left\{ \ln \left(\frac{r_j }{R_2} \right)
+ \ln\left[
\frac{2 \vartheta_1 \left(i \ln\left( \frac{R_2 }{r_j} \right), q\right)}
{i\, \vartheta_1'(0,q)}
\right]
\right\},
\label{annulus_phi}
\end{equation}
\medskip
\noindent
and
\begin{equation}
\begin{split}
V_{jk}&=\frac{k_j k_k \rho_* }{4\pi}\\
&\times\text{Re}\left[ \ln\left(
\frac{
\vartheta_1\left(
-\frac{i}{2} \ln\left(\frac{r_j r_k }{R_2^2}\right) + \frac{1}{2} \theta_{jk} , q
\right)
}{
\vartheta_1
\left(
-\frac{i}{2} \ln\left(\frac{r_j }{r_k }\right) + \frac{1}{2} \theta_{jk} , q
\right)
}
\right)\right]
\end{split}
\label{annulus_V}
\end{equation}
respectively, where $\vartheta_1(z,q)$ stands for the relevant Jacobi elliptic theta functions and $q=R_1/R_2$.

In the next sections, we focus on the scattering process of a single vortex dipole against the disk-like obstruction, i.e. we take $N_v=2$, and $N_1=+1$, $N_2=-1$. We characterize different dynamical regimes, within both the plane and the disk geometry.
Furthermore, for the sake of simplicity we restrict ourselves to the case of equal masses, $m_j=m_b N_b/2$, $j=1,2$.

Concerning the physical quantities, we take as the components two different atomic species, considering a binary mixture of
\ce{^{87}  Rb} (majority component $a$) and \ce{^{41}  K} (minority component $b$), experimentally realized for example as in Ref. \cite{Burchianti2018}. The scattering lengths are expressed in terms of the
Bohr radius $a_0$, with
$a_a=99a_0$, $a_b=65 a_0$, and $a_{ab}=163 a_0$ unless otherwise stated. Note that $a_{ab}$ is so to ensure the immiscibility condition, required for the presence of robust massive vortices, $g_{ab}>\sqrt{g_ag_b}$.
We take a trap of radius $R_2=50$ $\mu m$, a small central obstruction of radius $R_1=0.05 R_2$, 
and $d_z=0.04 R_2$. Finally, the particle number density $n_a=N_a/S$ is taken to be $n_a\simeq 1.28\times10^{13}$ $m^{-2}$, with $S$ the surface occupied by the superfluids, and $N_b= 10^2$ unless otherwise stated.

\section{Dipole scattering in a plane}
\label{sec:PLM_plane}

To investigate the scattering of a vortex-antivortex pair of massive vortices against a disk-like obstacle, we
begin considering a VA pair in an infinite plane, thereby excluding the effects of a confining potential. We make use of the PLM based on Lagrangian \eqref{eq:Lagr} considering the atomic species and physical quantities introduced in Section \ref{sec:PL_model}. 
The trajectories considered in the present discussion, illustrated in Figures \ref{fig:PL_symmetrical}-\ref{fig:massless},
feature the initial velocities $\dot{x}_1(0)=\dot{x}_2(0)=0$, $\dot{y}_1(0)=\dot{y}_2(0)\simeq  4.9\times 10^{-5}$ $m/s$ such that the initial dipole translation, referred to the motion of the dipole's center of mass, is aligned with the $y$-axis. 
In all the figures of the present and subsequent sections the black dots (white squares) indicate the initial (final) positions of the vortices, 
the circular obstacle is represented by the orange filled disk,
while the arrows mark the travel direction of the trajectories.
A preliminary analysis to this problem was inspired by the thesis in Ref. \cite{EnricoOrtu2024}.

Figure \ref{fig:PL_symmetrical} describes the scattering process of the dipole in the case of symmetric initial conditions $x_2(0)=-x_1 (0)$, $y_2(0)=y_1 (0)$. The dipole, after splitting, recovers its initial form and the vortex trajectories keep parallel to the $y$ axis, essentially recovering the dipole's initial velocity. 
After assuming the dipole dynamics of this symmetric case
as the reference case, we analyze a broad set of simulations 
characterized by asymmetrical initial conditions. These show how the scattering process essentially depends on two geometrical factors: 
the position of the dipole's center of mass with respect to the center of the obstacle (i.e. the origin $O$ of our reference frame),
measured by the impact parameter $b$, and the 
dipole length $d = |\vec d|$, i.e. the distance between the vortex initial positions. 
The parameter $b$ has modulus the distance of $O$ from the axis of the dipole vector 
${\vec d} ={\vec r}_2(0) -{\vec r}_1(0)$ at the initial configuration at $t=0$ (note that we choose ${\vec r}_1(0)$ and ${\vec r}_2(0)$ always with the same $x$-coordinate in our figures). The sign of $b$ is positive if the dipole axis lies on the right of the obstacle and negative otherwise.
The parameter $b$ is responsible for the emergence of an asymmetrical scattering characterized by a deflection angle $\phi$. 
The latter is defined as the deflection angle of the dipole translation line, after the scattering event, from the dipole translation line before the scattering event. 
Both $b$ and $\phi$ are  shown in Figure \ref{fig:PL_asymmetrical1} (right panel). The deflection angle, well visible in Figures
\ref{fig:PL_asymmetrical1}-\ref{fig:massless}, is zero in Fig. \ref{fig:PL_symmetrical} featuring the condition $b=0$ of full alignment.
Concerning the parameter $d$,
numerical simulations show how it influences the stability of the pair when this undergoes the scattering process. This aspect is discussed in the sequel.

It is important to remark that the trajectories considered in this paper do not show visible mass-related oscillations (see, for example, Refs. \cite{Richaud2021, Bellettini2023}), as we
on purpose choose examples of smooth enough trajectories at most exhibiting ripples of negligible size.
More specifically, we consider initial configurations such that, both before and after the scattering, the dipole length is almost constant and vortices proceed with essentially constant velocities along parallel trajectories that can be confused with straight lines.
We remark that this choice is justified by our primary interest in the scattering mechanism and the effects of the vortex mass on them, rather than on the small-oscillations properties of the system.

The combined changes of impact parameter $b$ and dipole length $d$ reveal an extremely rich phenomenology characterized by two main behaviors already observed by Griffin \textit{et al.} \cite{Barenghi2005}. 

The first behavior is the so-called go-around behavior (see Figs. \ref{fig:PL_symmetrical}, 
\ref{fig:PL_asymmetrical1} (left panel), and
\ref{fig:PL_asymmetrical2} (right panel)), which sees the dipole splitting at the obstacle and returning to its original length together deflecting its trajectory. In this process, the vortex that is closer to the obstacle, represented by the blue continued trajectory in the figures, circumnavigates the obstacle.
The second behavior is the so-called fly-by behavior (see Figs.
\ref{fig:PL_asymmetrical1} (right panel),
\ref{fig:PL_asymmetrical2} (left panel)), occurring when the two vortices essentially keep the same distance $d$ while they 
overtake the obstacle on one side, and the vortex (blue trajectory) does not circumnavigates it. 
As shown in Figs.
\ref{fig:PL_asymmetrical1}-\ref{fig:massless}, in both cases a deflection appears in relation to the asymmetrical initial positions with $b\ne 0$.

An interesting feature of the go-around behavior is the remarkable stability of the VA dipole: as can be seen in Figs.
\ref{fig:PL_symmetrical}, 
\ref{fig:PL_asymmetrical1} (left panel),
\ref{fig:PL_asymmetrical2} (right panel), and \ref{fig:massless}, the dipole, despite the splitting
effect occurring when the vortices overcome the disk, 
recombines after the separation. 
The dipole configuration gets significantly altered only 
close enough to the obstacle.
The latter introduces the forces of the virtual vortices and antivortices (representing the obstacle through the potentials described by equations \eqref{eq:V_plane} and \eqref{eq:phi_plane}) that disrupt the equilibrium state characterized, far enough from the obstacle, by an essentially constant dipole length and constant vortex velocities, depending on the interaction described by the potential $U_{12}$ (see Eq. \ref{eq:RV_plane}).

The stability of the dipole state in spite of scattering processes characterizes all the configurations involved in the VA dynamical evolutions according to the PLM. In other words, the two vortices always recombine, no matter how fast they approach to the obstacle or how far or close to it they are. 
To prove that the original dipole length, after bypassing the disk, is always restored, we resort to the total energy of the system 
$$
E = \sum_j \frac{m_j}{2} {\dot {\vec r}_j}^2 + \sum_j W_j + \sum_j \sum_{k \ne j} V_{kj},
$$
and to the observation that, for any configurations far enough from the obstacle  (and thus for both the initial and the final dipole evolution states), the kinetic energy term features vortex-velocity vectors whose final orientation can differ from the initial one (deflection effect) but whose moduli are the same. The kinetic energy is thus unchanged in the dipole-like asymptotic dynamical states. Since potentials ${\bar U}_{12} ( = {\bar U}_{21})$, $W_1$ and $W_2$, relevant to the obstacle, become negligible far enough from it (this condition in practice means $r_1, r_2 \ge 5\div 6\; R_1$), the remaining quantity characterizing the total VA energy is potential
${ U}_{12} (= { U}_{21})$ depending on $|{\vec r}_1 -{\vec r}_2|$. As a consequence, the energy conservation combined with the invariance of the kinetic energy implies the 
conservation of the dipole length $|{\vec r}_1 -{\vec r}_2|$.

Figs. \ref{fig:PL_asymmetrical1} and \ref{fig:PL_asymmetrical2} illustrate how the interplay between parameters $b$ and $d$ plays a crucial role in the transition between the two characteristic behaviors. 
Fig. \ref{fig:PL_asymmetrical1} (left panel)
shows how, when the antivortex (orange dashed curve) is far enough from the obstacle, the vortex (blue continuous curve) undergoes a go-around process because the virtual-antivortex action (mainly described by potential \eqref{eq:phi_plane}) prevails, thereby determining the vortex capture. This is equivalent to the formation of a virtual dipole formed by the vortex circumnavigating the obstacle and its virtual antivortex. 
On the other hand, upon decreasing the VA distance $d$ (and hence $b$) as in Fig. \ref{fig:PL_asymmetrical1} (right panel), the action of the virtual antivortex is inhibited and the VA pair exhibits the fly-by behavior.

Fig. \ref{fig:PL_asymmetrical2} illustrates a different effect governed by the same mechanism. In Fig. \ref{fig:PL_asymmetrical2} (left panel) the vortex (blue continuous) and the antivortex (orange dashed line) trajectories are 
closer than in Fig. \ref{fig:PL_asymmetrical1} (right panel). Not surprisingly, the further reduction of $d$ (stronger VA interaction described by potential $U_{12}$) keeps the fly-by behavior. However, as soon as the distance of the vortex from the obstacle is reduced, the virtual antivortex again triggers the dipole splitting determining the go-around effect of the vortex, as shown in Fig. \ref{fig:PL_asymmetrical2} (right panel). In this transition parameter $b$ is decreased while $d$ is increased.
It is interesting to observe how, during the go-around process, the vortex noticeably increases its velocity while the antivortex coherently undergoes a visible slow-down, to favor the dipole recombination immediately after the overcoming of the obstacle.

\subsection{Free-dipole massless subregime}
\label{sec:free_dipole}

An important feature that characterizes all the PLM simulations we show here is that the obstacle effects, dominated by the interaction with virtual vortices and responsible for the scattering, rapidly vanish at a finite distance from the obstacle.  
This feature allows us to simplify the VA-pair equations and then to highlight the presence of a well-defined dynamical subregime which justifies the dipole-like behavior of vortex motion far enough from the obstacle.
To this end, we rewrite
the VA-pair equations \eqref{eq:eq_rho} and
\eqref{eq:eq_theta} (previously expressed in the polar picture), in terms of Cartesian position vectors ${\vec r}_1 =(x_1, y_1)$ and ${\vec r}_2 =(x_2, y_2)$ with $k_1=-k_2=k$. The equations become
\begin{equation}
\begin{split}
&m_1  {\ddot {\vec r}_1} - \rho_* k {\vec e}_3 \wedge {\dot {\vec r}_1}  
= 
\frac{ \rho_*}{2 \pi} k^2
\Bigl [
{\vec F} ({\vec r}_1, {\vec r}_2 )
- \frac{ {\vec r}_1- {\vec r}_2 }{r_{12}^2}  
\Bigr ] ,
\\
&m_2  {\ddot {\vec r}_2} + \rho_* k {\vec e}_3 \wedge {\dot {\vec r}_2}  
=
\frac{\rho_* }{2\pi}  k^2  
\Bigl [  
{\vec F} ({\vec r}_2, {\vec r}_1 )
- \frac{ {\vec r}_2- {\vec r}_1 }{r_{12}^2}  \Bigr ] ,
\end{split}
\label{VAequation}
\end{equation}
where
$$
{\vec F} ({\vec r}_1, {\vec r}_2 )= 
\frac{ {r}^2_2 {\vec r}_1 - R_1^2 {\vec r}_2  }{ D }  
- \frac{ {\vec r}_1 }{ {r}^2_1 - R_1^2  } ,
\quad r_i= |{\vec r}_i|, \, 
i=1,2
$$
and
$$
D =r_1^2 {r}^2_2 + R_1^4
-2R_1^2 {\vec r}_1 \cdot {\vec r}_2, 
\quad r_{12}= |{\vec r}_1-{\vec r}_2|.
$$
For $r_i$ sufficiently larger than $R_1$ 
the field ${\vec F}$ becomes negligible and the obstacle no longer affects the dipole dynamics. In this case equations \eqref{VAequation} reduce to
\begin{equation}
\begin{split}
&m_1  {\ddot {\vec r}_1} - \rho_* k {\vec e}_3 \wedge {\dot {\vec r}_1}  
=
-\frac{ \rho_*}{2 \pi} k^2
\frac{ {\vec r}_1- {\vec r}_2 }{r_{12}^2} ,
\\
&m_2  {\ddot {\vec r}_2} + \rho_* k {\vec e}_3 \wedge {\dot {\vec r}_2} 
=
-\frac{\rho_* }{2\pi}  k^2  
\frac{ {\vec r}_2- {\vec r}_1 }{r_{12}^2} ,
\end{split}
\label{VAeqapp}
\end{equation}
and 
feature three constants of motion whose general form, for arbitrary $k_j$ and $m_j$, read
$$
J_x = \sum_j (\rho_* k_j x_j -m_j {\dot y}_j), \quad
J_y = \sum_j (\rho_* k_j y_j + m_j {\dot x}_j  ) ,
$$
$$
J_z = -\sum_j \Bigl [
m_j (x_j {\dot y}_j -y_j {\dot x}_j)) +\rho_* \frac{k_j}{2} r^2_j
\Bigr ].
$$
The conserved quantities $J_x$ and $J_y$ are easily found by summing the left-hand sides of Eqs. \eqref{VAeqapp}. This gives
$$
-
\sum_j \Bigl ( \rho_* k_j {\vec e}_3 \wedge { {\vec r}_j} - m_j  {\dot {\vec r}_j} \Bigr )= const
$$
whose vector components are $J_x$ and $J_y$. As observed above, Eqs. \eqref{VAeqapp} reveal the presence of a dynamical subregime. The latter is characterized by the constraints ${\ddot {\vec r}_1}= 0$ and ${\ddot {\vec r}_2}= 0$ at each time, thereby excluding acceleration effects. Note that this is equivalent to considering a zero-mass dynamics. We easily find that, in this regime, ${\dot {\vec r}_1}={\dot {\vec r}_2}$ while the time independence of ${\dot {\vec r}_j}$ is equivalent to the property that the quantities ${x_1}-{x_2}$ and ${y_1}-{y_2}$ are constant in time. They, in fact, are proportional to the constants of motion $J_x$ and $J_y$ for $m_j =0$ and $k_1=k=-k_2$
(see, e.g., \cite{Penna1999}).
Additionally, velocities ${\dot {\vec r}_1}$, ${\dot {\vec r}_2}$ are orthogonal to the dipole vector ${{\vec d}}= { {\vec r}_2}-{ {\vec r}_1} $. 
The subregime thus reproduces all the distinctive features of the motion of a massless free dipole, which excludes intervortex oscillations and behaves as a rigid entity. Note that, if initially ${\dot {\vec r}_2} \ne {\dot {\vec r}_1}$, the acceleration vectors must be reintroduced in Equations \eqref{VAeqapp}. 

An interesting characterization of the VA dynamics follows from reformulating Equations \eqref{VAeqapp} in a perturbative picture such that ${{\vec r}_j} = {{\vec s}_j}+{{\vec \eta}_j}$ with $|{{\vec \eta}_j}| << |{{\vec s}_j}|$. We find that the massless free-dipole subregime characterizing Equations
\eqref{VAeqapp} can be embedded in the zero-order dynamics of vectors ${{\vec s}_j}$, as the latter obey the same Equations \eqref{VAeqapp}
describing ${{\vec r}_j}$. The zero-order dynamics then features a free-dipole subregime with ${\dot {\vec s}_1}= {\dot {\vec s}_2}$ constant in time (the same observed in our figures far enough from the obstacle) where ${{\vec s}_j}$ supplies the dominating contribution to the vortex trajectories. In parallel, the first-order linear equations 
\begin{equation}
\begin{split}
&m_1  {\ddot {\vec \eta}_1} - \rho_* k {\vec e}_3 \wedge {\dot {\vec \eta}_1}
\simeq
-\frac{ \rho_*}{2 \pi} k^2
{\vec G}({\vec \eta}_{12}, {\vec s}_{12})  ,
\\
&m_2  {\ddot {\vec \eta}_2} + \rho_* k {\vec e}_3 \wedge {\dot {\vec \eta}_2} 
\simeq
-\frac{\rho_* }{2\pi}  k^2  
{\vec G}({\vec \eta}_{21}, {\vec s}_{21}) ,
\end{split}
\label{VAeq1ord}
\end{equation}
in which
$$
{\vec G}({\vec \eta}_{12}, {\vec s}_{12}) 
= \frac{{\vec \eta}_{12}}{s^2_{12}}
-2 \frac{
{\vec \eta}_{12} \cdot {\vec s}_{12} }{s^4_{12}} {\vec s}_{12},
$$
${\vec \eta}_{12} = {\vec \eta}_{1}-{\vec \eta}_{2}$, and
${\vec s}_{12} = {\vec s}_{1}-{\vec s}_{2}$,
determine the evolution of perturbations ${\vec \eta}_{j}$. Eqs. \eqref{VAeq1ord}, make it possible to describe configurations with slightly different initial velocities ${\dot {\vec r}_1} \ne {\dot {\vec r}_2}$
due to
${\dot {\vec \eta}_1} \ne {\dot {\vec \eta}_1}$ which entail accelerated motions such as, for example, intervortex oscillations.
This behavior is illustrated in Figure
\ref{fig:PL_asymmetrical_so}
where small but different initial velocities $\dot{x_1}(0)$ and $\dot{x_2}(0)$ are taken, together with a larger vortex mass characterized by the $b$-boson number
$N_b = 10^3$. Apart from the moderate deflection to the left shown by the dipole motion, the trajectories of the vortices exhibit small oscillations around their average positions $\vec{s_1} (t)$ and 
$\vec{s_2} (t)$, whose visibility was obtained thanks to the increasing of the mass of $b$-bosons.
Note that Eqs. \eqref{VAeq1ord} can lose their perturbative character for initial conditions involving a small initial intervortex distance
$|{\vec s}_{1} - {\vec s}_{2}|$ in the force field $\vec G$. 
This condition can cause an unreliable description of the dipole dynamics where the condition 
$|{\vec \eta}_j| << |{\vec s}_j|$ is violated.

\begin{figure}[ht]
\centering
\includegraphics[width=1\linewidth]{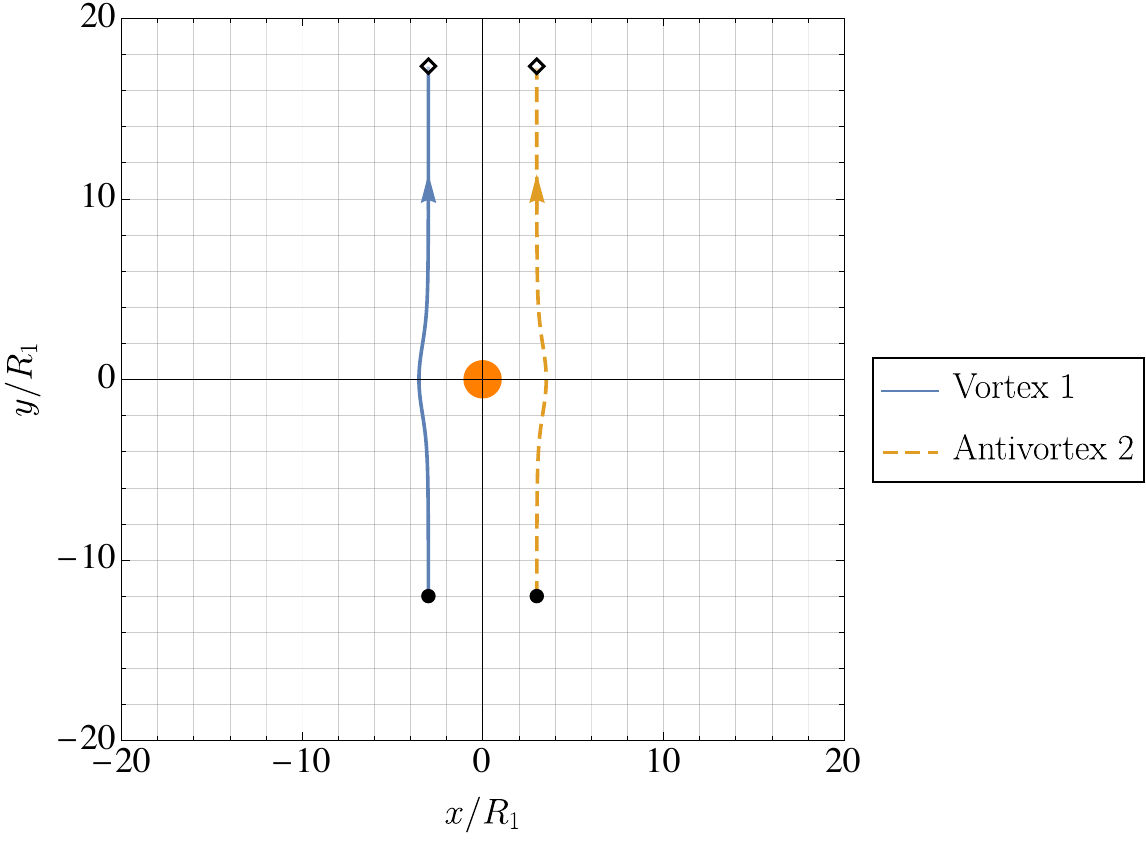}
\caption{Trajectory of a vortex dipole in an unbounded domain with a central obstacle: in case of symmetric initial positions (black dots) the recombination of the dipole occurs symmetrically with respect to both the $x$- and $y$-axes, as shown by the final positions (squares).The evolution time is $t=1.5$ $s$. 
} 
\label{fig:PL_symmetrical}  
\end{figure}

\begin{figure*}[ht]
\centering
\begin{minipage}{0.4\textwidth}
  \centering
  \includegraphics[width=1\linewidth]{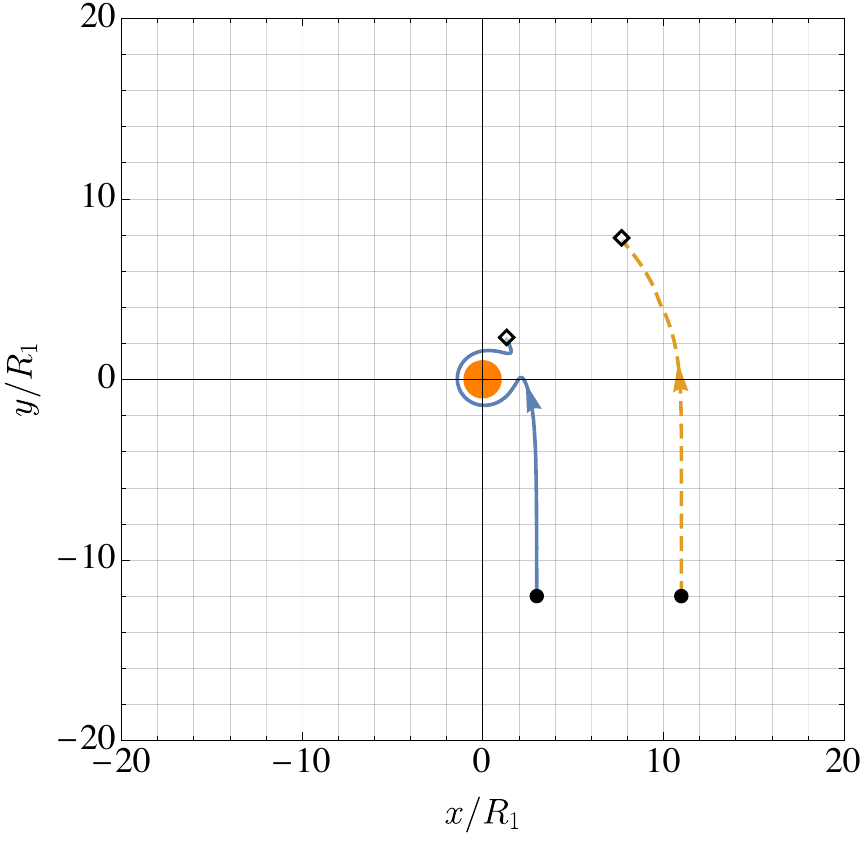}
\end{minipage}%
\begin{minipage}{0.53\textwidth}
  \centering
  \includegraphics[width=1\linewidth]{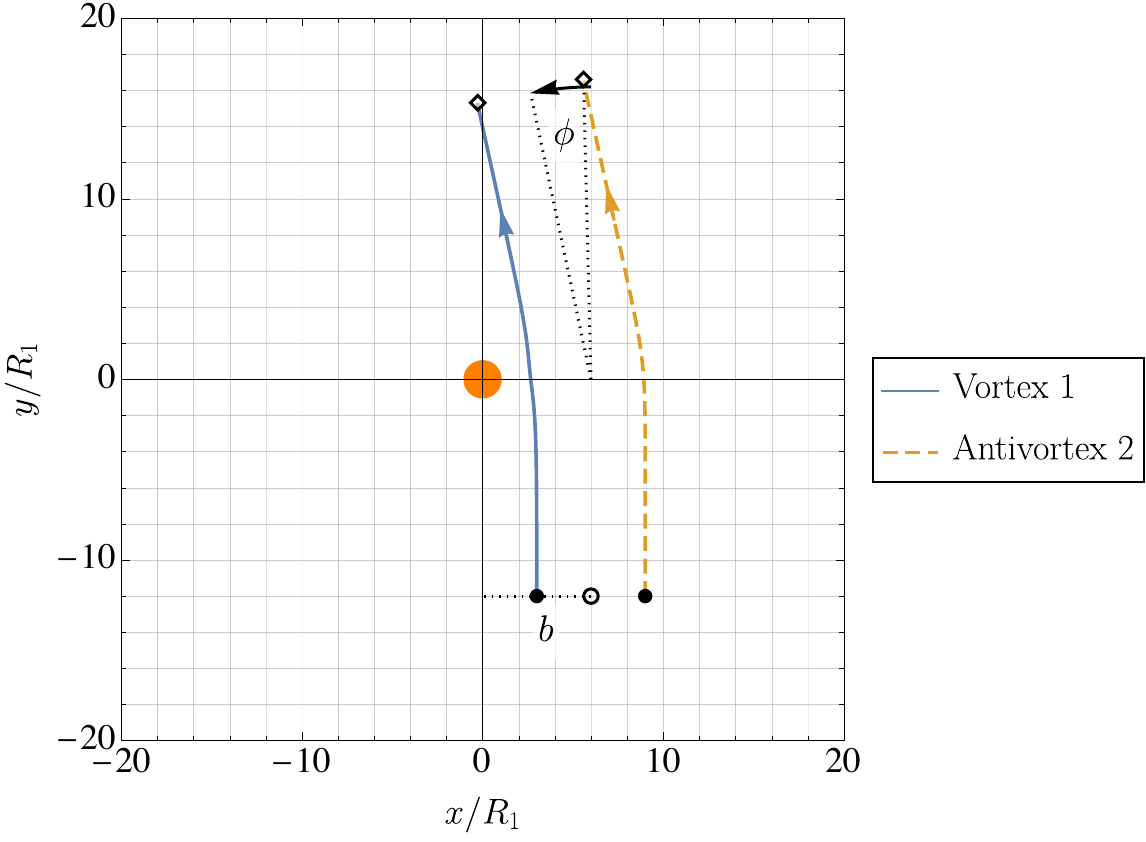}
\end{minipage}
\caption{The left (right) panel shows a go-around (fly-by) type of scattering, for a dipole in an unbounded domain with asymmetrical initial vortex positions. While the vortex is kept fixed, the initial position of the antivortex varies in the two panels. Note that a smaller dipole size corresponds to a higher velocity. Overall evolution time: $t =1.5$ $s$.}
\label{fig:PL_asymmetrical1}  
\end{figure*}

\begin{figure*}[ht]
\centering
\begin{minipage}{0.4\textwidth}
  \centering
  \includegraphics[width=1\linewidth]{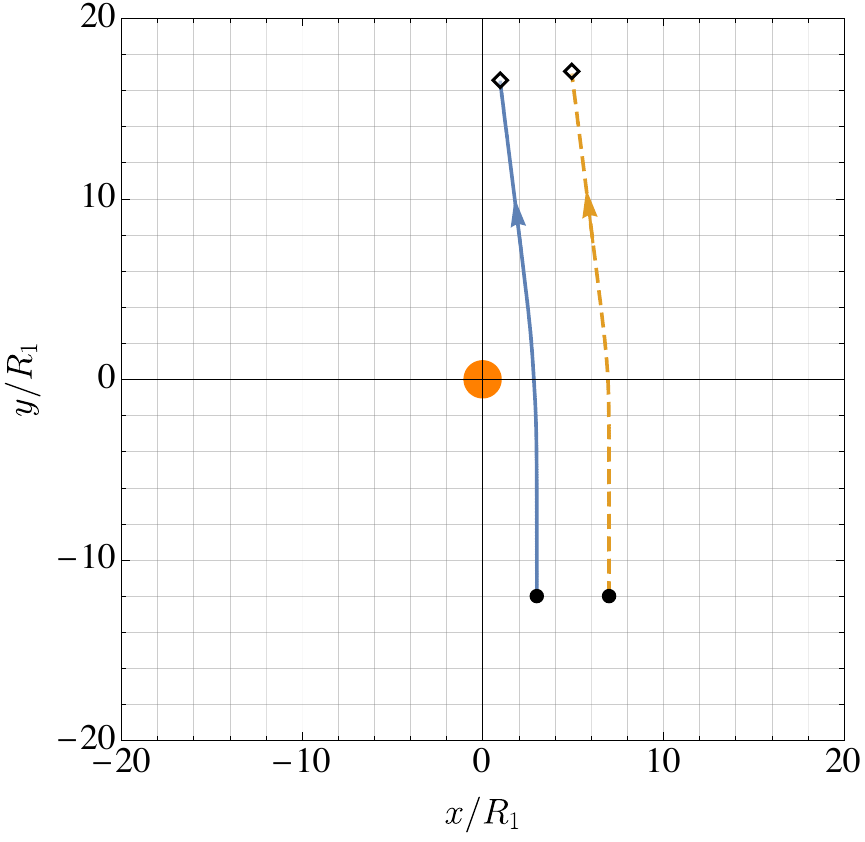}
\end{minipage}%
\begin{minipage}{0.53\textwidth}
  \centering
  \includegraphics[width=1\linewidth]{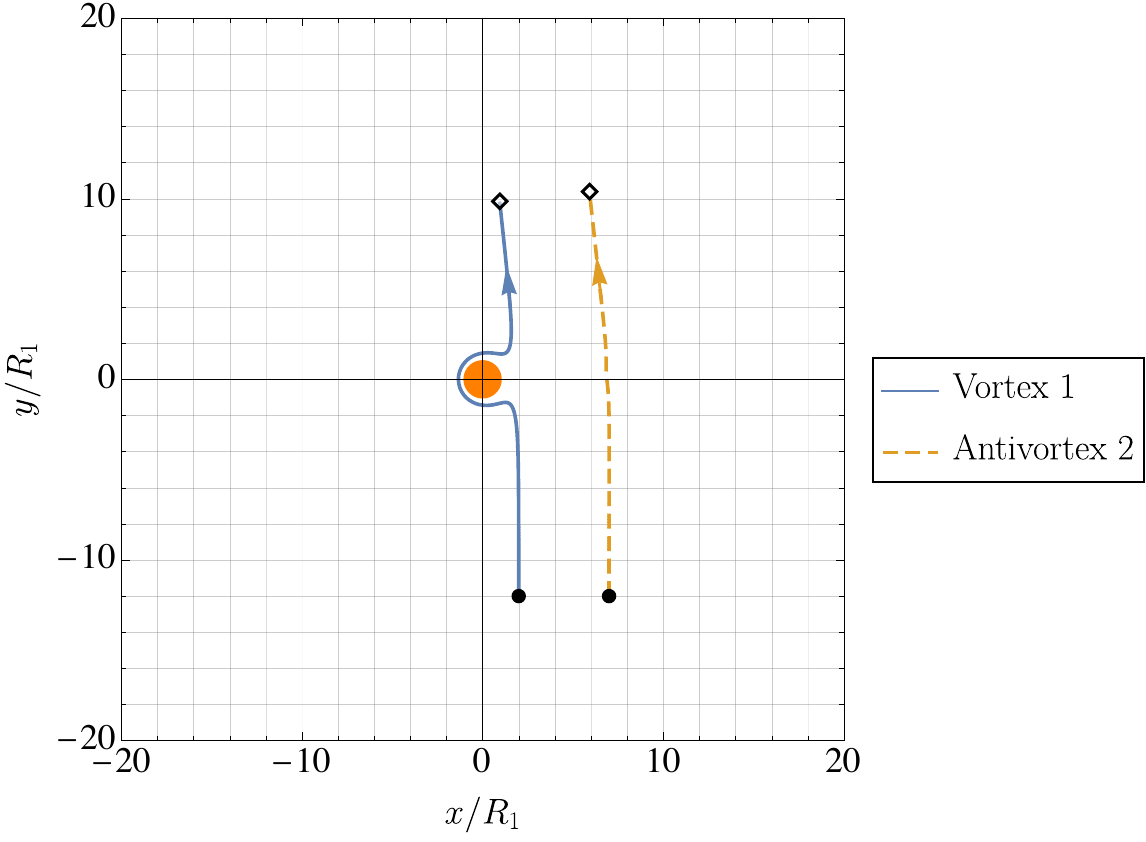}
\end{minipage}
\caption{The left (right) panel presents the fly-by (go-around) of a vortex dipole in its scattering against an obstacle in an unbounded domain. In the left panel, the vortex
is kept at the same initial position of the vortex in the right panel of Fig. \ref{fig:PL_asymmetrical1}, while the antivortex is placed closer to the obstacle. 
This choice still preserves the fly-by process, thus increasing the dipole's overall velocity. On the other hand, keeping now fixed the antivortex, and moving the vortex closer to the obstacle, induces a go-around type of scattering, which involves an asymmetric recombination of the dipole (right panel). Evolution time: $t=1$ $s$.}
\label{fig:PL_asymmetrical2}  
\end{figure*}

\begin{figure*}[ht]
\centering
\begin{minipage}{0.4\textwidth}
  \centering
  \includegraphics[width=1\linewidth]{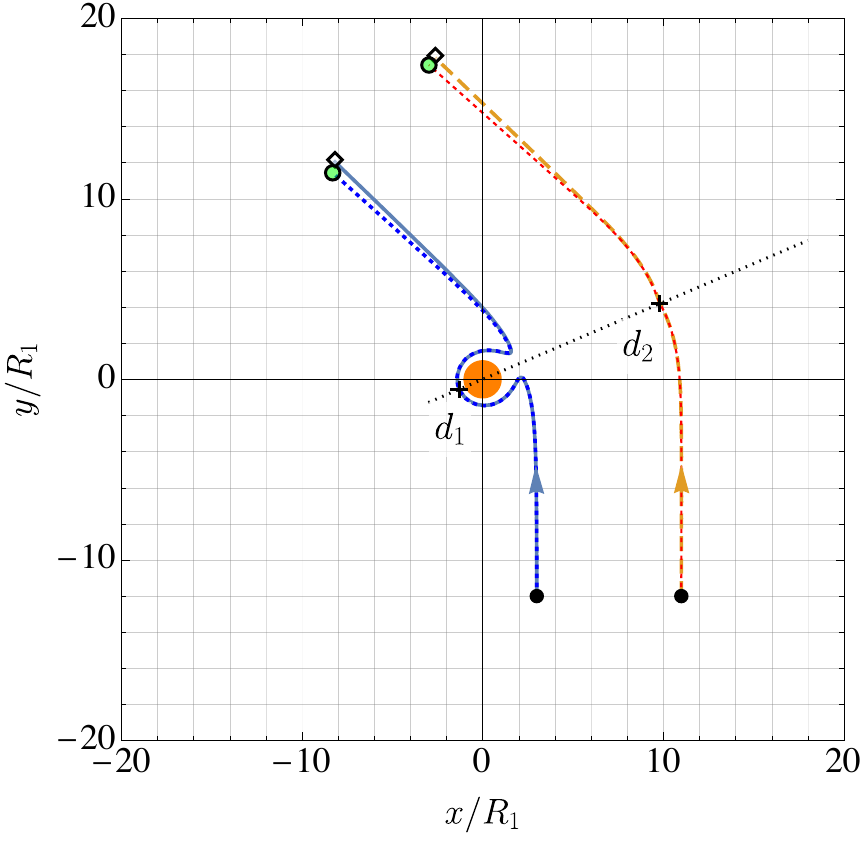}
\end{minipage}%
\begin{minipage}{0.56\textwidth}
  \centering
  \includegraphics[width=1\linewidth]{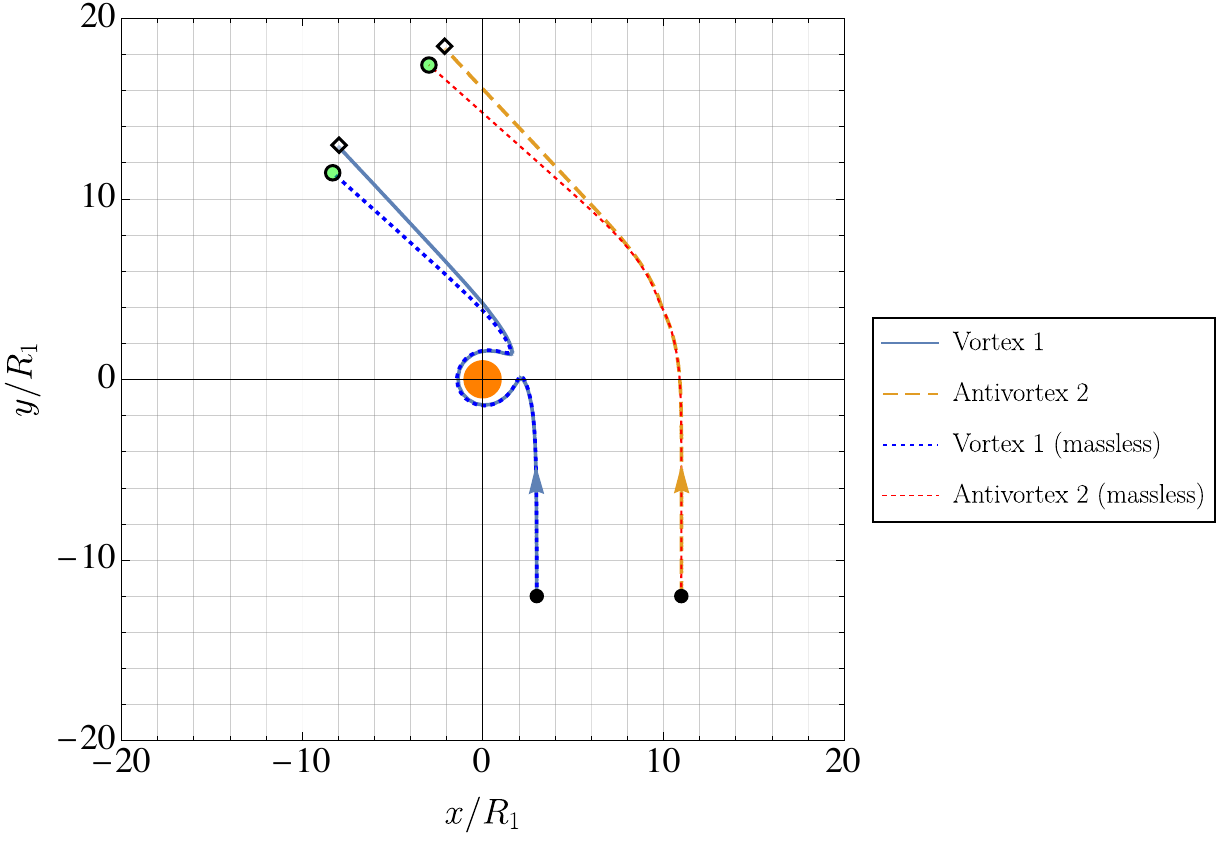}
\end{minipage}
\caption{Comparison of the scattering dynamics of a massive dipole (continuous lines) and a massless dipole (dotted lines) against a disk in a go-around scenario and in an unbounded domain. Left panel sees trajectories for the usual core boson number of $N_b=10^2$, where the slight variation in the dipole translation line after the recombination is attributable to the nonlinear interaction between the obstacle and the core masses. Apart from this, the agreement in the trajectories of the two dynamics shows how, within the asymptotic regimes, it is possible to reduce the massive dynamics to its massless counterpart. The right panel shows greater differences due to an increase vortex mass ($N_b=2.5\times 10^2$): the deflection angle is increased due to the stronger centrifugal effect caused by
the heavier cores. Evolution time: $t = 2.5$ $s$.} 
\label{fig:massless}  
\end{figure*}

Without performing a systematic stability analysis of perturbation Equations \eqref{VAeq1ord} (this is beyond the scope of the present paper), numerical simulations confirmed that the oscillations (the ripples described by ${\vec \eta}_{j}$) around the massless-case  trajectories described by ${\vec s}_{j}$ maintain a small size during the time evolution. 
In particular, we remark that the trajectories of Figs. \ref{fig:PL_symmetrical}-\ref{fig:massless}
feature a purely dipole-like behavior and do not involve visible oscillations due to the choice of the initial conditions. 

The previous perturbative treatment can be easily extended to Equations \eqref{VAequation} in the region, close to the obstacle, where the scattering takes place and confirms the predominance of the massless-dipole behavior. In particular, Fig.
\ref{fig:massless} (left panel) shows that the vortex trajectories of the massless case are essentially indistinguishable from that of the massive case, not only during the initial evolution, but also during the interaction with the obstacle. Only after the recombination the vortex masses manifest their presence inducing a small but evident deviation of $\phi$ from the massless case caused by the centrifugal force. The role of the masses in the scattering process is further evidenced by Fig. \ref{fig:massless} (right panel) which shows how a larger vortex mass increases such deviation from the massless dipole behavior. This observation highlights the non-secondary effect of the vortex masses during the scattering event.

Concerning our simulations, it is worth noting how the excessive proximity of one vortex to the disk can affect our numerical approach producing  divergent contributions in the PLM equations \eqref{VAequation} (or in their polar version \eqref{eq:eq_theta}). These are responsible for non-physical behaviors such as the collision of one vortex with the obstacle. 
It is important to emphasize that while the PLM breaks down in such limit cases, the GPE are still able to provide a physical solution as it will be
discussed in Sec. \ref{sec:GPE}.

\subsection{Deflection angle vs impact parameter}

\begin{figure}
    \centering
    \includegraphics[width=1\linewidth]{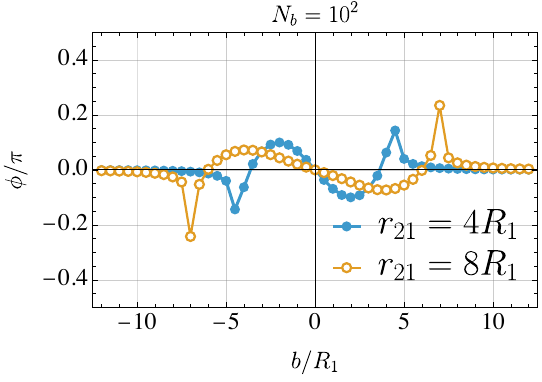}
    \caption{Deflection angle $\phi$ of the dipole trajectory, after a scattering event in an unbounded domain, as a function of the impact parameter $b$, for two different values of the dipole size $r_{21}$, and for $N_b=10^2$. The initial velocities are, for the case at $r_{21}=4R_1$ ($r_{21}=8R_1$), $\dot{x}_1(0)=\dot{x}_2(0)=0$, $\dot{y}_1(0)=\dot{y}_2(0) \simeq 7.3\times  10^{-5}$ $m/s$ ($\dot{y}_1(0)=\dot{y}_2(0) \simeq 3.7\times  10^{-5}$ $m/s$).}
    \label{fig:theta_h_100}
\end{figure}

Figure \ref{fig:theta_h_100}, inspired by Ref. \cite{Griffin2017}, gives some quantitative results on the dipole scattering in the plane for given, relatively low, vortex masses. In the figure, we plot the scattering angle $\phi$ with respect to the impact parameter 
$b$, where both variables are represented by the sketch in Fig. \ref{fig:PL_asymmetrical1}. 
The information that can be inferred from Fig. \ref{fig:theta_h_100} are significative: for two instances of the vortices initial distance $r_{12}=r_2(0)-r_1(0)$,
we find that the plot of $\phi(b)$ highlights the transition between the fly-by and go-around scattering behaviors, and that the go-around region can feature both positive and negative scattering angles. The plot is, as expected, symmetric with respect to the origin, and, starting from large
negative values of $b$, it features: \textit{i)} a region where $\phi$ starts close to zero
and the dipole is in the fly-by region, \textit{ii)} a region, marked by an initial negative peak of $\phi$, where one vortex
is captured by the obstacle and the dipole enters the go-around process. Within this region, at lower negative values of $b$, there is a change in the sign of $\phi$, which becomes positive: this transition is explained by the 
interplay of the real- and virtual-vortex interactions.
Figure
\ref{fig:scattering_sequence} illustrates instances of dipole trajectories for values of $b$ spanning throughout all the scattering behaviors. It is interesting to examine, in Fig.
\ref{fig:theta_h_100} the effect of a larger dipole size (i.e. $r_{12}$). It is visible how, as a direct effect of a weaker interaction between the (real) vortices the capturing by the obstacle occurs ``earlier'', i.e. at larger values of $|b|$, and the $\phi$ peak and antipeak are more pronounced (larger trajectory deflection). In all cases, at large enough $|b|$, the interaction with the obstacle becomes negligible and $\phi$ approaches zero, at the case of $b=0$ is the symmetric go-around scattering case.
Note that the cases shown in Fig. \ref{fig:theta_h_100} do not feature any trajectories corresponding to the break-down of the PLM (gray zone), whereas, depending on the system's parameters, it can happen that not all the values of $b$ are allowed as a solution of the PLM. 
The figure points out a technique for controlling the translation deflection of a vortex-dipole via the artificial insertion, in the system, of obstacles.
Interestingly in Ref. \cite{Griffin2017}, Griffin \textit{et al.} produced a figure analogous to Fig. \ref{fig:theta_h_100} for the case of massless vortices, basing on the GPE,
and found the same trends. This result further validates the point-like models.

\section{Analytic characterization of the dipole scattering in a plane}
\label{sec:scatt}

In this section, based on the dynamical Equations
\eqref{VAequation}
for a vortex dipole with $k = k_1 = -k_2 >0$, 
we study the scattering process of a VA pair from a disk-like obstruction
within an analytical perspective. We again focus on the simplest possible geometry by excluding the influence of any external confining potentials. 
More specifically, 
we highlight a characterization of the dipole scattering in terms of simple but significant geometric parameters, such as the minimum distances of vortices from the disk center. Also, we determine the critical condition for which the dipole splitting takes place.

The discussion in Section \ref{sec:PLM_plane}
and the vortex dynamics illustrated in Figs. 
\ref{fig:PL_symmetrical}-\ref{fig:PL_asymmetrical2}
show how, far from the disk, the VA pair moves as a massless free dipole, in which vortices essentially exhibit the same velocity vectors. On the other hand, in the proximity of the disk, the dipole can undergo a simple deviation which does not alter the dipole structure (fly-by scattering) or, more interestingly, can be momentarily split showing the capture of one of the vortices by the disk (go-around scattering). In the latter case, vortex $2$ considerably slows down waiting for the recombination with vortex $1$, which circumnavigates the disk.

In general, as suggested by Figs. \ref{fig:PL_asymmetrical1} and \ref{fig:PL_asymmetrical2}, the scattering of the dipole features a simple geometry characterized by the presence of a symmetry axis describing the invariance under reflection 
of vortex trajectories. The axis, illustrated in Fig.
\ref{fig:massless} (left panel), corresponds to the instant $t=t_0$ 
in which vortices reach the minimum distance from the disk center. In this configuration both the vortex positions and the disk center lie on a common straight line corresponding to the symmetry axis. Numerical simulations confirm the property that, independently from the initial conditions, the vortex trajectories describing the scattering always exhibit this axis.

The equations of motion for a VA pair written
in terms of polar coordinates $r_j$ and $\theta_j$ are described by formulas \eqref{eqr1}-\eqref{eqth2} derived in Appendix \ref{A1}. We define the angles $\theta_j$ in ${\vec r}_j = r_j (\cos \theta_j, \sin \theta_j)$ with respect to the Cartesian frame where the horizontal axis is assumed to  coincide with the symmetry axis ($\theta_j = 0$). Also, 
as in Figs. \ref{fig:PL_asymmetrical1} and \ref{fig:PL_asymmetrical2}, vortex $1$ (vortex $2$) represents the vortex closer (farther) to the disk. The configuration where vortex $1$ encircles the disk (dipole splitting) is, at time $t=t_0$, characterized by
\begin{equation}
\theta_1 = \pi, \quad \theta_2 = 0, \quad {\dot r}_2 = {\dot r}_1 = 0,
\label{case1}
\end{equation}
while the case involving a fly-by  scattering is described by
\begin{equation}
\theta_1 = \theta_2 = 0, \quad {\dot r}_2 = {\dot r}_1 = 0.
\label{case2}
\end{equation}
In both cases the angular-component equations imply that ${\ddot \theta}_1 = {\ddot \theta}_2 =0$ (see Appendix \ref{A1}) while the radial-component equations take a form apt to investigate the scattering process when considering configurations \eqref{case1} and \eqref{case2}. We denote the time-dependent variables $r_j$, $\theta_j$, ${\dot r}_j$ and ${\dot \theta}_j$ at the time $t_0$ with the new symbols
$$
d_1 = r_1(t_0) , \,\, d_2 = r_2(t_0), \,\, 
\omega_1 = {\dot \theta}_1 (t_0), \,\,
\omega_2 = {\dot \theta}_2 (t_0),
$$
defining the radial distance from the disk center and the relevant angular velocities (see the representation of $d_1$ and $d_2$ in the left panel of Fig. \ref{fig:massless}).
\medskip

\noindent
{\it Case 1: Go-around scattering}. 
The application of conditions \eqref{case1}, characterizing the scattering with dipole splitting, reduces the radial-component equations \eqref{eqr1} and \eqref{eqr2}
to the form 
$$
- m_j d_j \omega_j^2 
-
 (-1)^j \rho_* k d_j \omega_j =
$$
\begin{equation}
\frac{\rho_* k^2}{2 \pi} \left [ \frac{d_k}{d_1d_2 +R_1^2} -\frac{d_j}{d_j^2 -R_1^2} 
-\frac{ 1}{d_1+d_2} \right ], \,\, j \ne k,
\label{radeq12c1}
\end{equation}
with $j,k = 1,2$. 
The solution of the previous equations gives the angular velocities
\begin{equation}
\omega_1^{\pm}
=
\frac{\rho_* k }{ 2m_1 } 
\left [ 1 \pm
\sqrt{ 1+ \frac{2m_1}{\pi \rho_* d_1} A(d_1,d_2) }
\right ],
\label{dphi1}
\end{equation}
\begin{equation}
\omega_2^{\pm} 
=
\frac{\rho_* k }{ 2m_2 } 
\left [ -1 \pm \sqrt{ 1+ \frac{2m_2}{\pi \rho_* d_2} A(d_2,d_1) }
\right ],
\label{dphi2}
\end{equation}
with
$$
A(r_1,r_2) = 
\frac{ 1}{r_1+r_2} -\frac{r_2}{r_1r_2 +R_1^2} +\frac{r_1}{r_1^2 -R_1^2} .
$$
Since vortex $1$ encircles the disk, $\omega_1$ and $\omega_2$ are expected to be respectively negative and positive. This condition is only satisfied by considering the angular velocities $\omega^-_1$ and $\omega^+_2$. Appendix \ref{A1} explores the limiting case of $d_1 \simeq R_1$, with a generic $d_2 > R_1$, 
which allows to considerably simplify the formulas for $\omega^-_1$ and $\omega^+_2$. In this case, the latter are expressed by Eqs. \eqref{dph1A} and \eqref{dph2A}. 
In particular, if the effective masses associated to the annuli around the disk $M_1= \pi \rho_* ( d_1^2 -R_1^2)$ and $M_2= \pi \rho_* ( d_2^2 -R_1^2)$ are sufficiently larger than the vortex masses $m_1$ and $m_2$, respectively, Eqs. \eqref{dph1A} and \eqref{dph2A} reduce to the meaningful expressions
\begin{equation}
\omega^-_1 
\simeq
\frac{-k }{ 2\pi ( d_1^2 -R_1^2) } <0, \,\,\,
\omega^+_2 
\simeq
\frac{k }{ 2\pi ( d_2^2 -R_1^2) }>0. 
\label{dph12app}
\end{equation}
It is important to note that, while $d_2$ sufficiently larger than $R_1$ is typically assumed in the cases of physical interest we consider, the condition $d_1 \simeq R_1$ (reflecting the  proximity of vortex $1$ to the disk) could become nonphysical if $d_1$ is too close to $R_1$. This limiting situation is excluded from the present analysis.
 
Remarkably, Eqs. \eqref{dph12app} highlight how vortices $1$ and $2$, in this intermediate time at step $t=t_0$ of the scattering process, {\it i)} exhibit angular velocities that are independent from the vortex masses, {\it ii)} exclude the contributions of the vortex-antivortex interaction (screened by the obstacle) and {\it iii)} are governed only 
by the force fields of potentials \eqref{eq:phi_plane} associated to the corresponding virtual vortices. Note that the fact that $d_1$ is sufficiently smaller than $d_2$ in Eq. \eqref{dph12app} entails the inequality $|w_1^- | >> w_2^+$ implying in turn that vortex $1$ rapidly encircles the obstacle while vortex $2$ \textit{slows down} to wait for the dipole final recombination.
\medskip

\noindent
{\it Case 2: Fly-by scattering}. 
Implementing the conditions \eqref{case2} (no dipole splitting) reduces the radial-component Eqs.
\eqref{eqr1} and \eqref{eqr2} to the form 
$$
- m_j d_j \Omega_j^2 
-
(-1)^j \rho_* k d_j \Omega_j =
$$
\begin{equation}
\frac{\rho_* k^2}{2 \pi} \left [ \frac{d_k}{d_1d_2 - R_1^2} -\frac{d_j}{d_j^2 -R_1^2} 
+\frac{ (-1)^j }{d_1-d_2} \right ], \,\, j \ne n,
\label{radeq12c2}
\end{equation}
where $j,k = 1,2$,
and the angular velocities $\omega_j$ were denoted, in the current case, by $\Omega_j$.
The sign-dependent factors $(-1)^j$ are related to the fact that, unlike case $1$ for which ${\vec r}_1 \cdot {\vec r}_2 = -r_1 r_2$, in the current case the positions of vortices are such that ${\vec r}_1 \cdot {\vec r}_2 = r_1 r_2$. The solution of the previous equations gives the angular velocities
\begin{equation}
\Omega_1^{\pm} 
=
\frac{\rho_* k }{ 2m_1 } 
\left [ 1 \pm
\sqrt{ 1 - \frac{2m_1}{\pi \rho_* d_1} B(d_1,d_2) }
\right ],
\label{Omega1}
\end{equation}
\begin{equation}
\Omega_2^{\pm} 
=
\frac{\rho_* k }{ 2m_2 } 
\left [ -1 \pm \sqrt{ 1 - \frac{2m_2}{\pi \rho_* d_2} B(d_2,d_1) }
\right ].
\label{Omega2}
\end{equation}
with
$$
B(r_1,r_2) = 
\frac{ 1}{r_2-r_1} +\frac{r_2}{r_1r_2 -R_1^2} -\frac{r_1}{r_1^2 -R_1^2} .
$$
In this case, the physically meaningful choices in Eqs. \eqref{Omega1} and \eqref{Omega2} correspond to the angular velocities $\Omega_1^-$ and $\Omega_2^+$ which, in general, exhibit positive values. 
We find evidence of the validity of such expressions through the limiting case $d_1, d_2 >> R_1$ describing a dipole far from the disk. Based on formulas \eqref{dph1B} and \eqref{dph2B} derived from \eqref{Omega1} and \eqref{Omega2}, one gets

\begin{equation}
{\Omega}^-_1  
\simeq \frac{ k}{2\pi d_1 d_{21}} >0 , \quad
{\Omega}^+_2  
\simeq \frac{ k}{2\pi d_2 d_{21}} > 0,
\label{Om12}
\end{equation}
with $d_{21} = d_2-d_1$. These angular velocities are physically meaningful because they reproduce the expected velocities $v_1 = v_2$, with $v_j = d_j \Omega_j$, of vortices $1$ and $2$ far from the obstacle, reflecting the dipole-like behavior of the VA pair.

\begin{figure}
    \centering
    \includegraphics[width=1\linewidth]{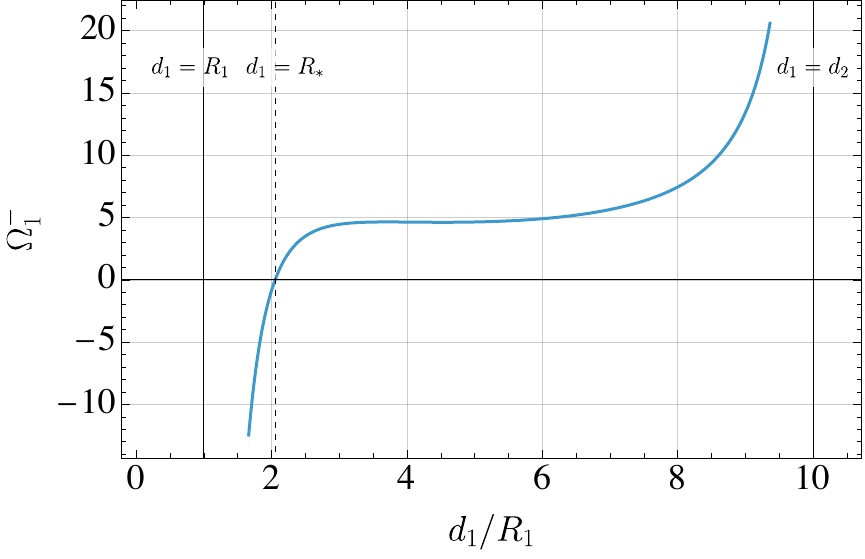}
    \caption{Angular velocity $\Omega_1^-$ as in Eq. \eqref{Omega1}, in the case of a fly-by type of scattering. Through its non-physical solutions for $\Omega_1$ (i.e. $\Omega_1<0$), the equation indirectly 
    catches the transition, at $d_1\le R_*$, to a scattering behavior where the vortex $1$ is captured by the obstacle (go-around process). This scenario is possible at $d_2$ large enough (in this case $d_2=10 R_1$). In this case, we took $N_b=5\times 10^2$, $m_1=m_2=N_b m_b/2$.
    }
    \label{fig:configs_transition}
\end{figure}

Anyway, the interesting information of the current case emerges when we analyze more carefully the behavior of $\Omega^-_1$ at given $d_2$, with $R_1 < d_2$, for a generic $d_1 \in [R_1, d_2]$. We can identify a critical value of $d_1$ for which $\Omega_1$ becomes negative. The condition $\Omega^-_1 \ge 0$ crops up from the inequality 
\eqref{2ineq} in Appendix \ref{A1}, and amounts to setting $0 \le B(d_1, d_2)$. A simple calculation shows that the latter formula is equivalent to 
\begin{equation}
d_2 \le d_1 \left [1 + \frac{d_1^2-R_1^2 }{ 2R_1^2} \left ( 1 + \sqrt{ 1+ \frac{ 4R_1^2}{d^2_1}} \right ) \right ].
\label{crit1}
\end{equation}
Based on the expression of $B(d_1, d_2)$, we evince the trend of $\Omega_1^-$ at fixed $d_2$ as a function of $d_1$ (see Eq. \eqref{Omega1}), illustrated in
Figure
\ref{fig:configs_transition}.
The figure 
shows a
critical value $R_* >R_1$ at which the function $\Omega_1(B(d_1, d_2))$ (as well as $B(d_1, d_2)$) intercepts the horizontal axis ($B(d_1, d_2)=0$). In other words, the angular velocity $\Omega^-_1$ becomes negative for $R_1 < d_1 < R_*$. This critical value defines the 
point at which the break down of the fly-by behavior takes place.
The inequality \eqref{2ineq} also provides the upper limit $R_1' \simeq d_2$, which excludes complex nonphysical values of $\Omega_1^-$.

\section{Dipole scattering in a closed domain}
\label{sec:PLM_annulus}

An intermediate step towards an accurate comparison with the Gross-Pitaevskii equations (see Section \ref{sec:GPE}) passes through the introduction of a boundary at finite distance so that the massive dipole is still scattered by a disk-like obstacle 
but confined into a circular trap of radius $R_2= 20\times R_1 = 50$ $\mu m$ (displayed by a continuous black line). This extra trapping term is included in the PLM via Eqs.
\eqref{annulus_V} and \eqref{annulus_phi} by means of the virtual-vortex method.
The simulations presented in the present section have been performed with initial velocities aligned with the $y$-axis: $\dot{x}_1(0)=\dot{x}_2(0)=0$, $\dot{y}_1(0)=\dot{y}_2(0)\simeq  3.7 \times 10^{-5}$ $m/s$.

\begin{figure*}[ht]
\centering
\begin{minipage}{0.4\textwidth}
  \centering
  \includegraphics[width=1\linewidth]{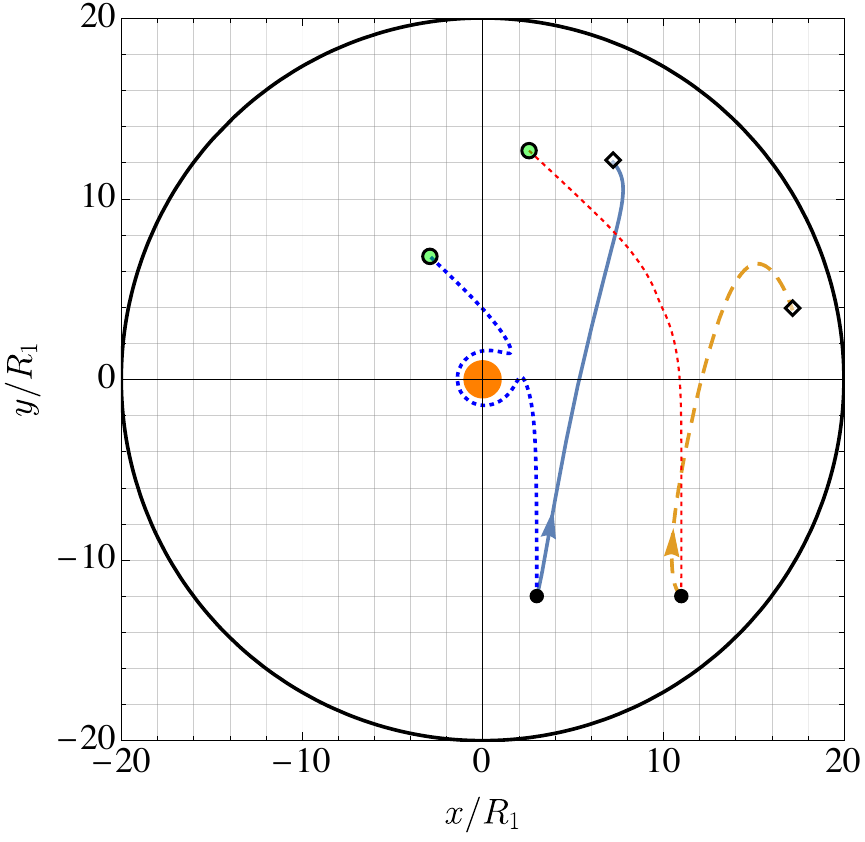}
\end{minipage}%
\begin{minipage}{0.55\textwidth}
  \centering
  \includegraphics[width=1\linewidth]{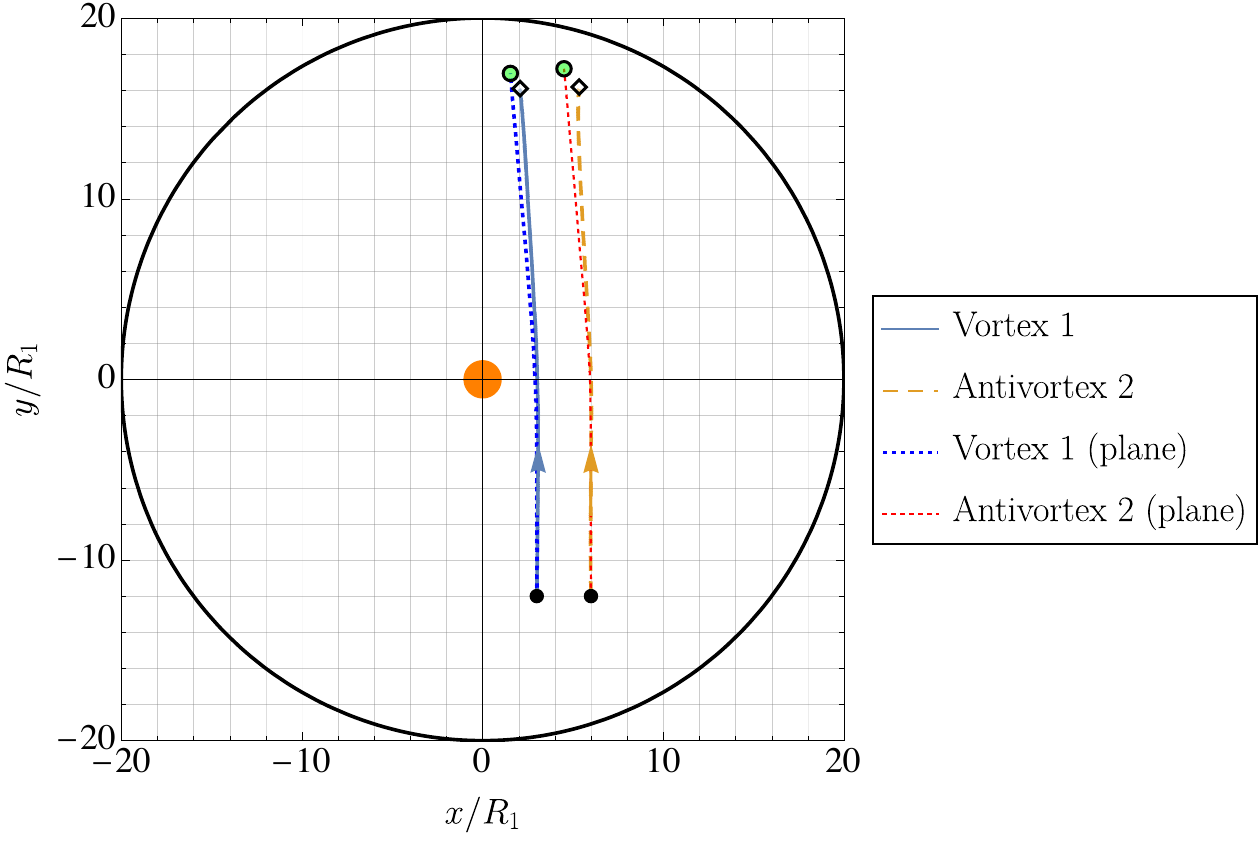}
\end{minipage}
\caption{Comparison of the scattering dynamics of a massive dipole in an unbounded domain (dotted lines) vs a confined domain (continuous and dashed lines).
The left panel highlights the remarkable effects of the external boundary on the scattering process: whereas in the plane geometry a go-around process takes place, the presence of the external disk-like boundary induces a fly-by behavior.
In the right panel, conversely, the deviation in the two cases mostly appears close to the external boundary, when the dipole splits.
Evolution times: $2$ $s$ (left panel) and $0.75$ $s$ (right panel).  }
\label{fig:annulusvsplane}  
\end{figure*}

While both the go-around and fly-by still describe the two main
scattering behaviors of a dipole in the current geometry, 
Figure \ref{fig:annulusvsplane} (right panel) clearly shows that the external boundary can significantly affect the scattering process.
Specifically, the figure presents a comparison of scattering processes within the plane and the confined geometry. 
In the left panel, whereas the plane geometry features a go-around type of scattering, the confined geometry
induces a fly-by scattering of the dipole (at equal initial conditions). 
Conversely, in the right panel of Fig. \ref{fig:annulusvsplane},
the distance of the dipole from the two boundaries make the effects of the external confinement essentially negligible while the scattering event against the obstacle takes place.

The second major effect of the external boundary is to induce the splitting of the dipole as the latter collides with it.
This splitting is
followed by the
two vortices running along the external boundary in opposite directions (due to the dominating effect of virtual vortices). 
Remarkably, despite this temporary breaking of the dipole due to boundary effects, the dipole then recombines as a consequence of the circular geometry of the confining boundary.

\begin{figure}[ht]
\centering
\includegraphics[width=1\linewidth]{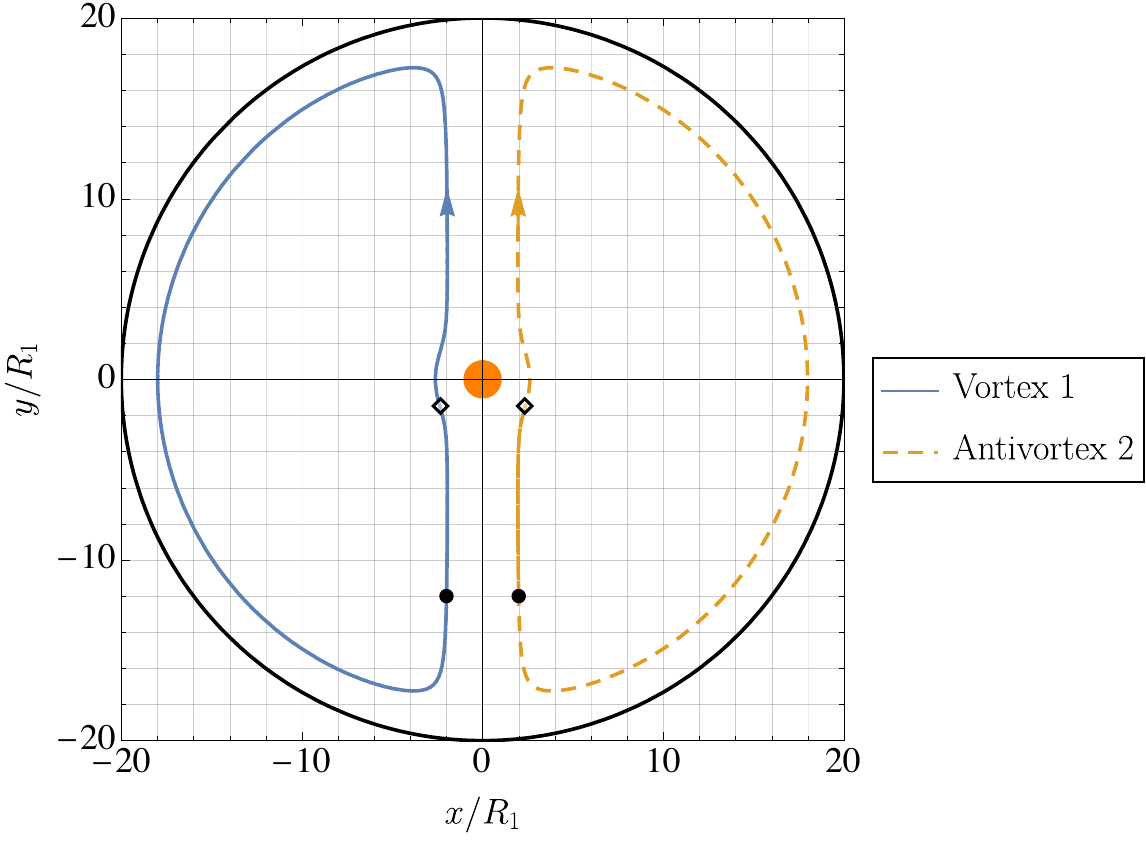}
\caption{Symmetrical VA scattering in a confined domain: as in the unbounded-domain case, symmetrical initial conditions lead to a perfectly symmetrical recombination. The presence of a boundary at finite distance results in the splitting of the dipole close to the border followed by its (symmetrical) recombination, ending in closed trajectories. The simulation time is of $3.75$ $s$.}
\label{fig:longtime}  
\end{figure}

Figure \ref{fig:longtime} shows the case of trajectories that are symmetric with respect to the $y$-axis due to the symmetry of the VA initial conditions.

\begin{figure*}[ht]
\centering
\begin{minipage}{0.4\textwidth}
  \centering
  \includegraphics[width=1\linewidth]{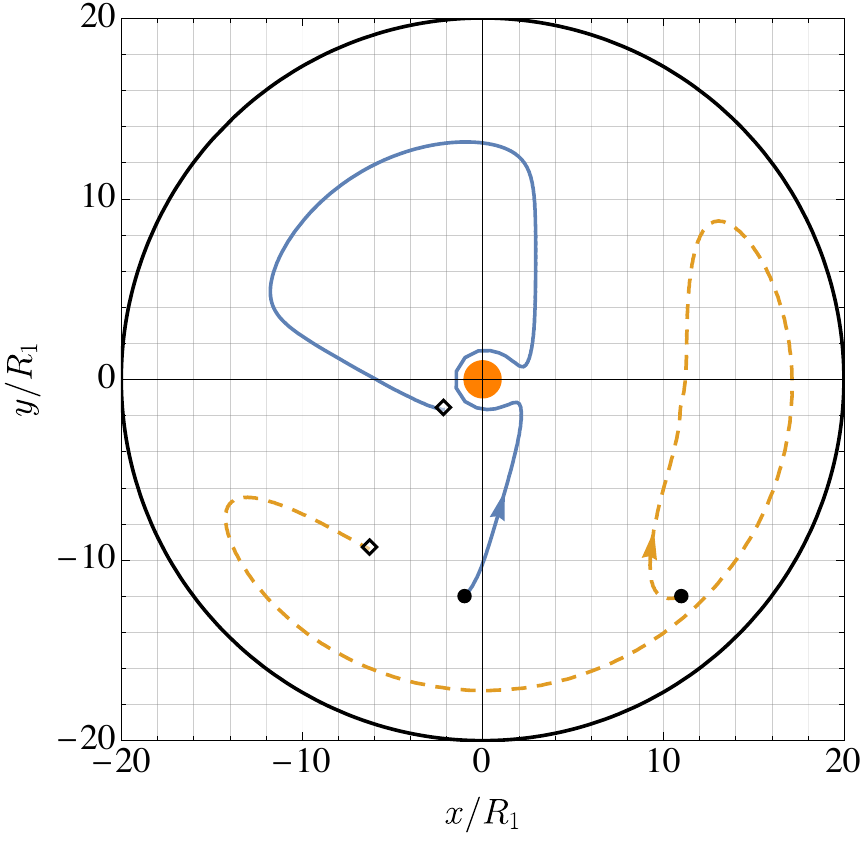}
\end{minipage}%
\begin{minipage}{0.51\textwidth}
  \centering
  \includegraphics[width=1\linewidth]{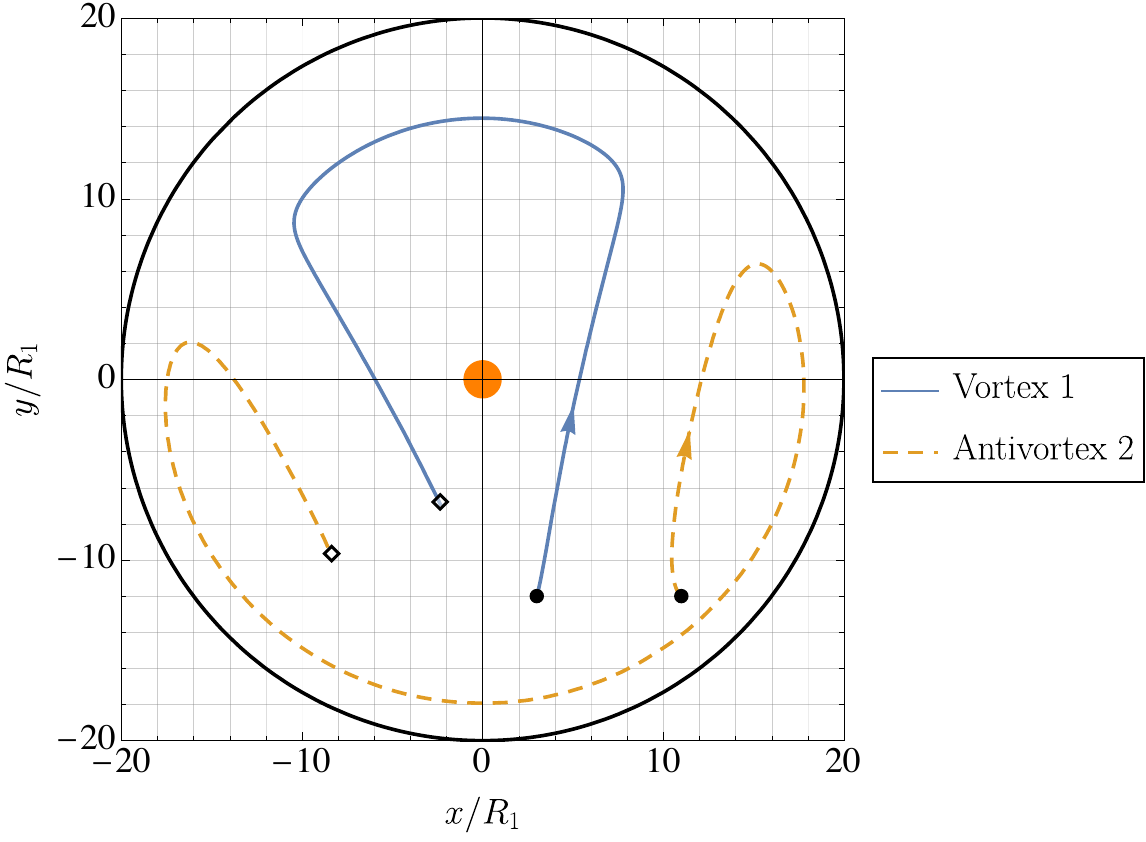}
\end{minipage}
\caption{Go-around (left panel) and fly-by (right panel) behaviors in a confined domain at varying initial vortex positions.
The evolution times are $8$ $s$ (left panel) and $6$ $s$ (right panel). }
\label{fig:ann_long_time}  
\end{figure*}

In the left and right panels of Figure \ref{fig:ann_long_time},
depicting a go-around and a fly-by scenario, respectively, it is possible to observe how the antivortex (orange dashed line) exhibits a motion along the external boundary with a constant velocity that is larger than the (constant) velocity of the vortex (blue continuous line). This is the consequence of the greater proximity of the antivortex to the outer boundary. One should in fact recall that, in this situation, the antivortex forms a virtual dipole with its virtual vortex (representing the boundary), thus moving with constant velocity along the boundary. The same feature characterizes the boundary motion of the vortex, which combines with its virtual antivortex. 
However, in the right panel, both the vortex and the antivortex move along the boundary faster than their counterparts in the left panel due to their greater proximity to the boundary. For this reason the vortex and antivortex trajectories in the right panel describe circular arcs whose length is larger.
It is worth noting how, right after the recombination, the VA pair reproduces, up to a rotation, the initial conditions. This implies that the VA motion replicates thus manifesting a cyclic behavior. 
This mechanism characterizes the trajectories of both the left and right panels. 

Longer simulation times
confirm that this splitting and recombination patterns always appear rotated of some angle, possibly different every time. This behavior strongly reminds a quasi-periodic regime in which vortices never travel on previous trajectories, thus covering all the possible points in the domain for large times.
As for the simulations of Sec. \ref{sec:PLM_plane}, the system exhibits some degree of sensibility to the initial conditions. Slight variations on the vortices’ initial positions may strongly influence the resulting dynamics especially when close to the transition zone between the two scattering behaviors. 
Despite the higher complexity arising from the introduction of a second (confining) boundary,
no
chaotic behavior seems to be detectable, as the VA nonlinear dynamics always preserves ordered trajectories and the dipole stability.

\section{Gross-Pitaevskii equations}
\label{sec:GPE}

Two coupled Gross-Pitaevskii equations
well describe, at a mean-field level, the ground state and the dynamical evolution of the condensates wavefunctions $\psi_a$ and $\psi_b$ \cite{Pitaevskii2016}. 
The GPEs, coupled through the inter-component repulsion $g_{ab}$, read

\begin{equation}
\begin{aligned}
    i\hbar\, \dot  \psi_a   &=  -\frac{\hbar^2}{2m_a} \nabla^2 \psi_a \\
    &+ \left(V_{ \text{ext}}(\vec{r}) + \frac{g_{ab}}{d_z} |\psi_b |^2 + \frac{g_{a}}{d_z}  |\psi_a |^2  \right) \psi_a  , \\
    i\hbar\, \dot \psi_b  &= -\frac{\hbar^2}{2m_b} \nabla^2 \psi_b \\
    &+ \left( V_{ \text{ext}}(\vec{r}) + \frac{g_{ab}}{d_z}  |\psi_a |^2+ \frac{g_{b}}{d_z}  |\psi_b |^2  \right)  \psi_b ,
\end{aligned}
\label{eq:GPEs}
\end{equation}

with the parameters as described in Section \ref{sec:PL_model}, and derivable from the Lagrangian \eqref{eq:L_GP} as the relevant (field) Euler-Lagrange equations.
Our simulations are based on an imaginary-time algorithm for the production of the initial state, represented on a $256\times256$ spatial grid, and on the Runge-Kutta method for the resolution of the time-discretized GPEs. The vortex positions are extracted via a space average weighted by $(1-|\psi_a|^2/n_{a,\text max})$, with $n_{a,\text max}$ the maximum value of the field $|\psi_a|^2$.

\subsection{Comparison with the PLM}
\label{sec:GPE_PLM}

Remarkably, despite the more complex geometry of the confined domain, the numerical simulations confirm the presence of the two
scattering behaviors predicted by the PLM
(i.e. the go-around and the fly-by, see Sec. \ref{sec:PLM_annulus}),  
and are qualitatively in agreement with it. The 
simplest case is obtained when the VA pair has symmetric initial conditions.
In this case, a go-around scenario takes place.
In general, when comparing the PLM and GPEs results, we observe that the quantitative discrepancies arise due to stronger boundary effects in the GPEs simulations than those predicted by the PLM.

\begin{figure*}[ht]
\centering
\begin{minipage}{0.4\textwidth}
  \centering
  \includegraphics[width=1\linewidth]{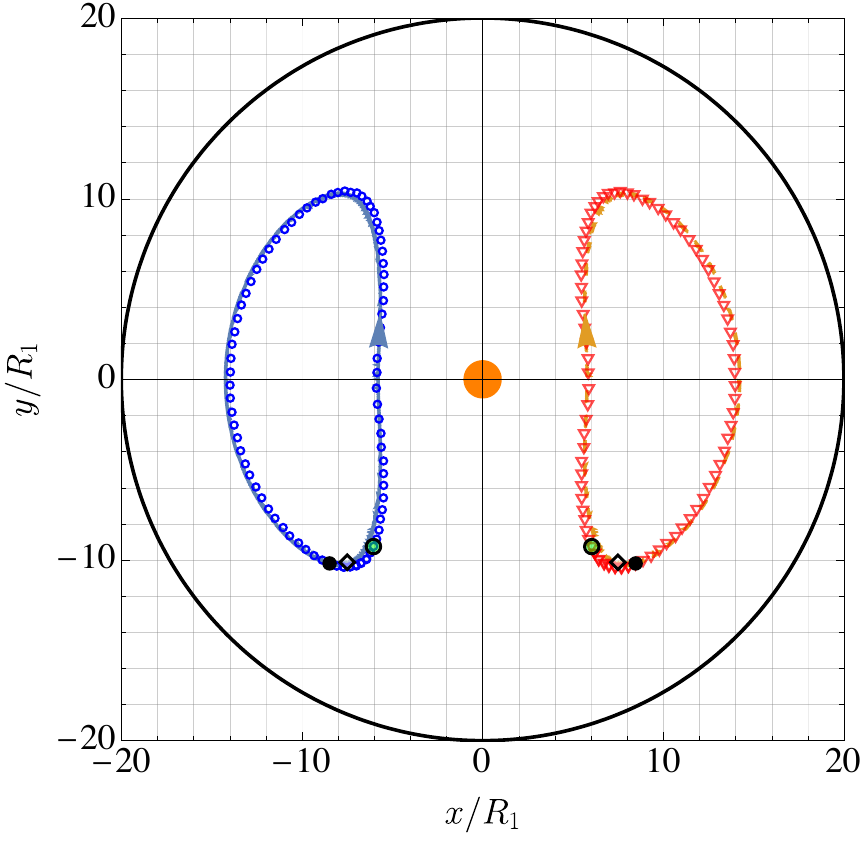}
\end{minipage}%
\begin{minipage}{0.576\textwidth}
  \centering
  \includegraphics[width=1\linewidth]{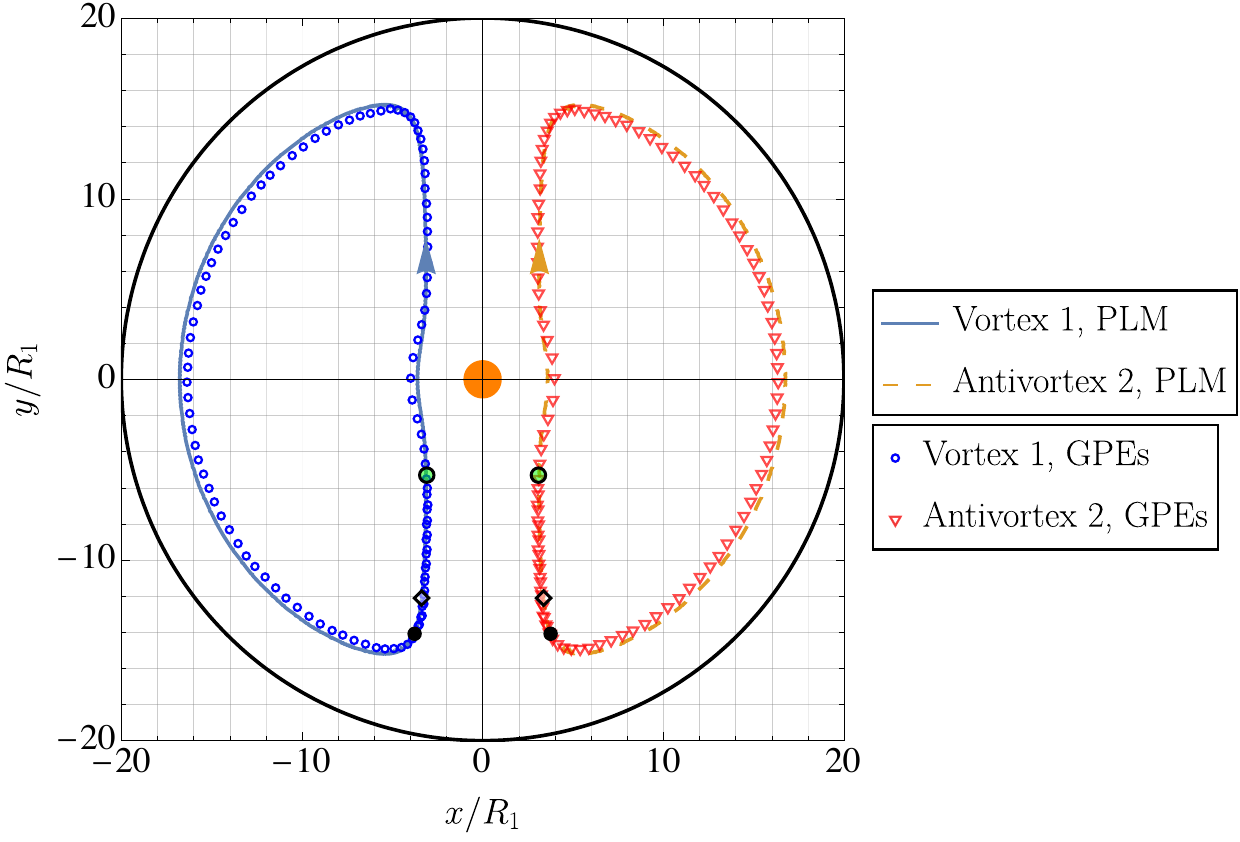}
\end{minipage}
\caption{Trajectory of a vortex dipole in a disk-like confining trap with a central obstacle: comparison of the GPEs results versus the PLM for an example of go-around scattering. The left panel features trajectories that stay further from the boundaries, in comparison to the right panel. The white diamonds (green circles) represent the final positions of the vortices according to the PLM (GPEs). In both the cases, $N_b=5\times10^2$, and 
the initial conditions on the $x$-velocities are $\dot{x}_1(0)=\dot{x}_2(0)=0$, $\dot{y}_1(0)=\dot{y}_2(0)$, with $\dot{y}_1(0)\simeq 1.7\times  10^{-5}$ $m/s$ and $\dot{y}_1(0)\simeq 3.8\times 10^{-5}$ $m/s$, respectively for the left and right panel.
The evolution time in the left (right) panel is of $8.8$ ($5.4$) $s$.
}
\label{fig:symm_trajs_comp}
\end{figure*}

{\it Symmetric initial conditions}. 
Figure \ref{fig:symm_trajs_comp} illustrates two symmetric dipole trajectories, comparing the PLM results and the numerical simulations.
The two vortices, initially placed at the points highlighted by the black dots, translate, split and recombine in the dipole.
In the two panels we report two cases at different distances from the boundaries.
In general, we see a very good quantitative agreement between the PLM and the GPEs, where the GPEs trajectories deviate from the model as a consequence of an earlier boundary effect. While far from the boundaries the dipole tends to translate, as the boundary effect kicks in, the dipole splits and the vortices move along a curved trajectory. As visible in the figure, the dipole splitting in the GPEs trajectories precedes that predicted by the PLM.

\begin{figure*}[ht]
\centering
\begin{minipage}{0.4\textwidth}
  \centering
  \includegraphics[width=1\linewidth]{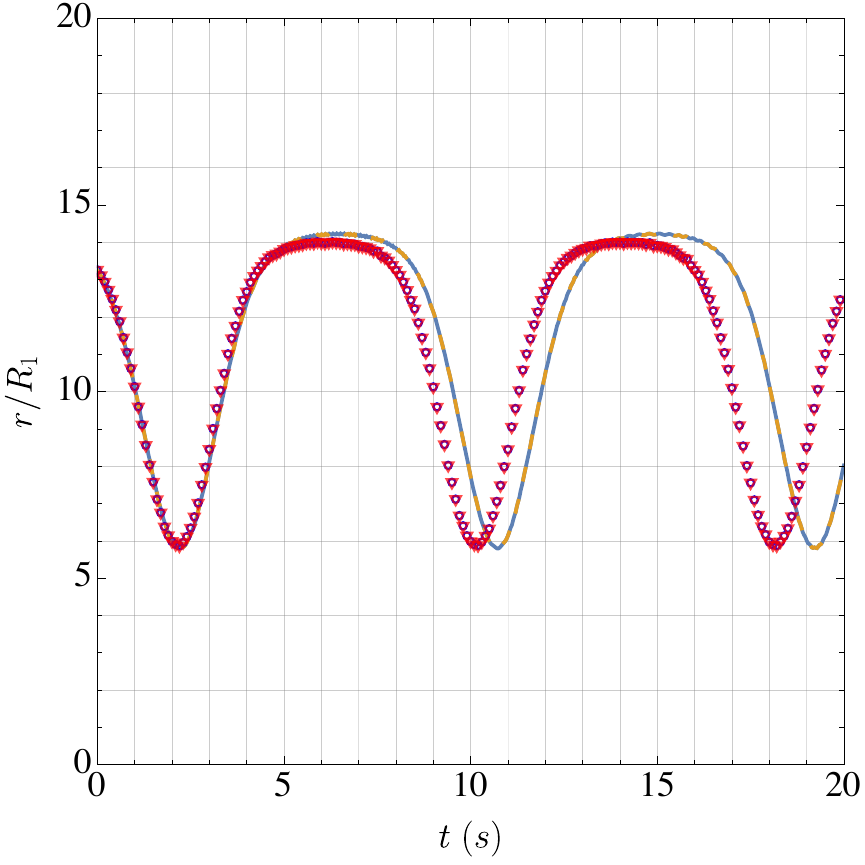}
\end{minipage}%
\begin{minipage}{0.576\textwidth}
  \centering
  \includegraphics[width=1\linewidth]{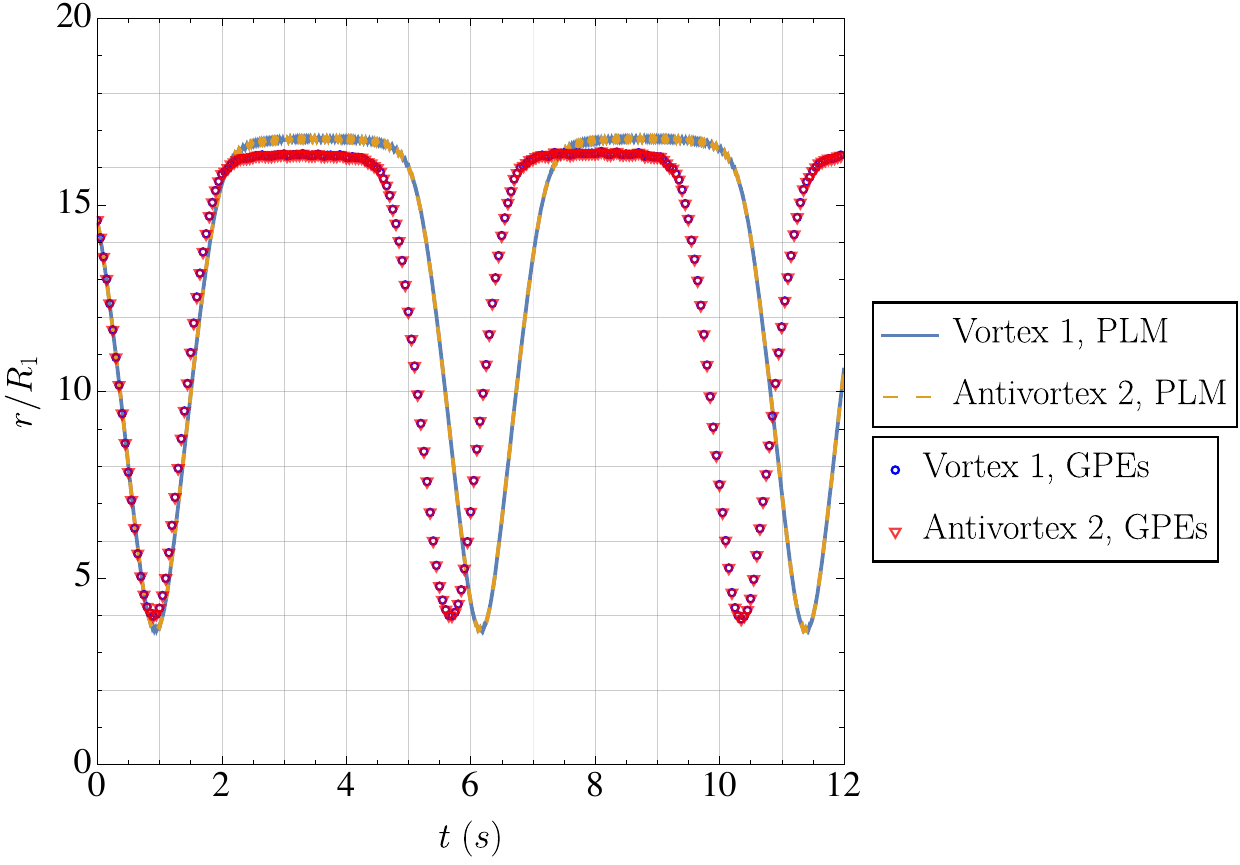}
\end{minipage}
\caption{Comparison between the vortices radial positions $r_1$ and $r_2$ as predicted by the PLM versus the results of the GPEs simulations, for a centrally symmetric dipole motion. The system in the left (right) panel is the same of the left (right) panel of Fig. \ref{fig:symm_trajs_comp}.}
\label{fig:symm_radii_comp}
\end{figure*}

This difference in the sensitiveness to the boundary is even more clearly visible by comparing the vortices radial positions displayed in Figure \ref{fig:symm_radii_comp}, relevant to the trajectories of Fig. \ref{fig:symm_trajs_comp}.
Here the plateau-like regions represent tracts of approximately circular motion, induced by the boundaries. The onset of such motion occurs slightly earlier in the GPEs simulations with respect to the PLM, inducing an increasing shift between the vortex positions in the mean-field description as compared to the PLM.
The comparison of the radial coordinates $r_1$ and $r_2$ corroborates the very good agreement between PLM and GPEs up to the boundary effects, in that it shows that the Gross-Pitaevskii (GP) dynamics is almost periodic over a long time, and the vortex trajectories roll up on themselves, as predicted by the PLM.
The shift between PLM and mean-field dynamics, induced by the boundary, can also be inferred in Fig. \ref{fig:symm_trajs_comp} by comparing the final vortex positions according to the GPEs, highlighted by the green circles, and the final positions as predicted by the PLM, represented by white diamonds.

Another feature worth remarking, is the average larger velocity of the vortices in the right panel of Fig. \ref{fig:symm_trajs_comp} with respect to the left panel, due to the smaller dipole size and to the lower
distance of the vortex (antivortex) from its virtual antivortex (vortex) in the proximity of the boundary.

\begin{figure*}[ht]
\centering
\begin{minipage}{0.4\textwidth}
  \centering
  \includegraphics[width=1\linewidth]{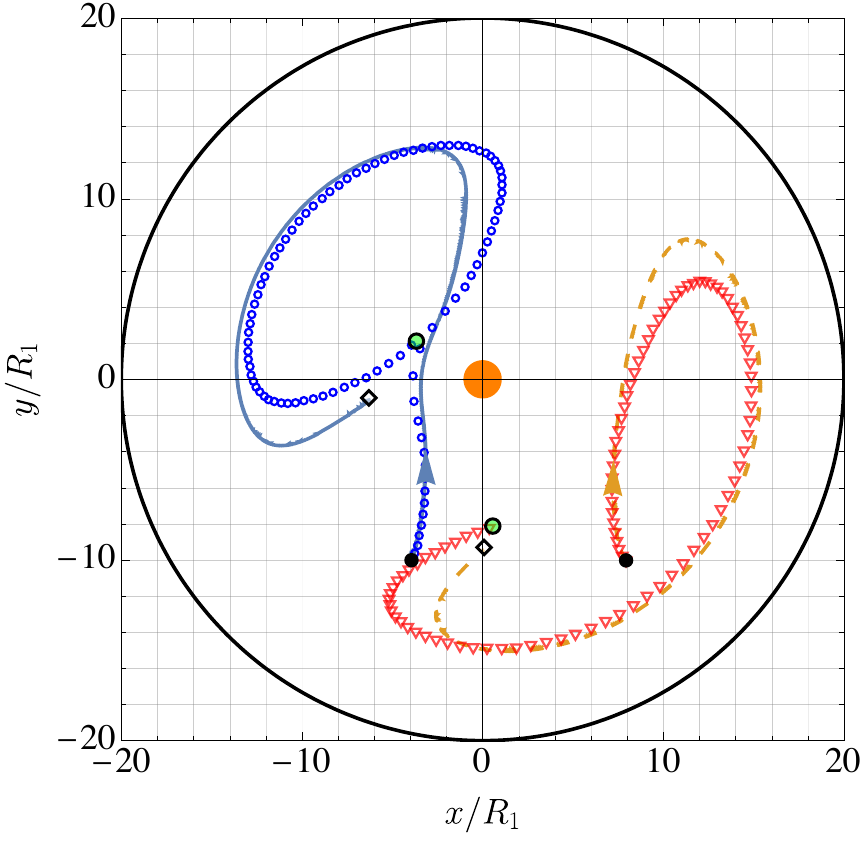}
\end{minipage}%
\begin{minipage}{0.576\textwidth}
  \centering
  \includegraphics[width=1\linewidth]{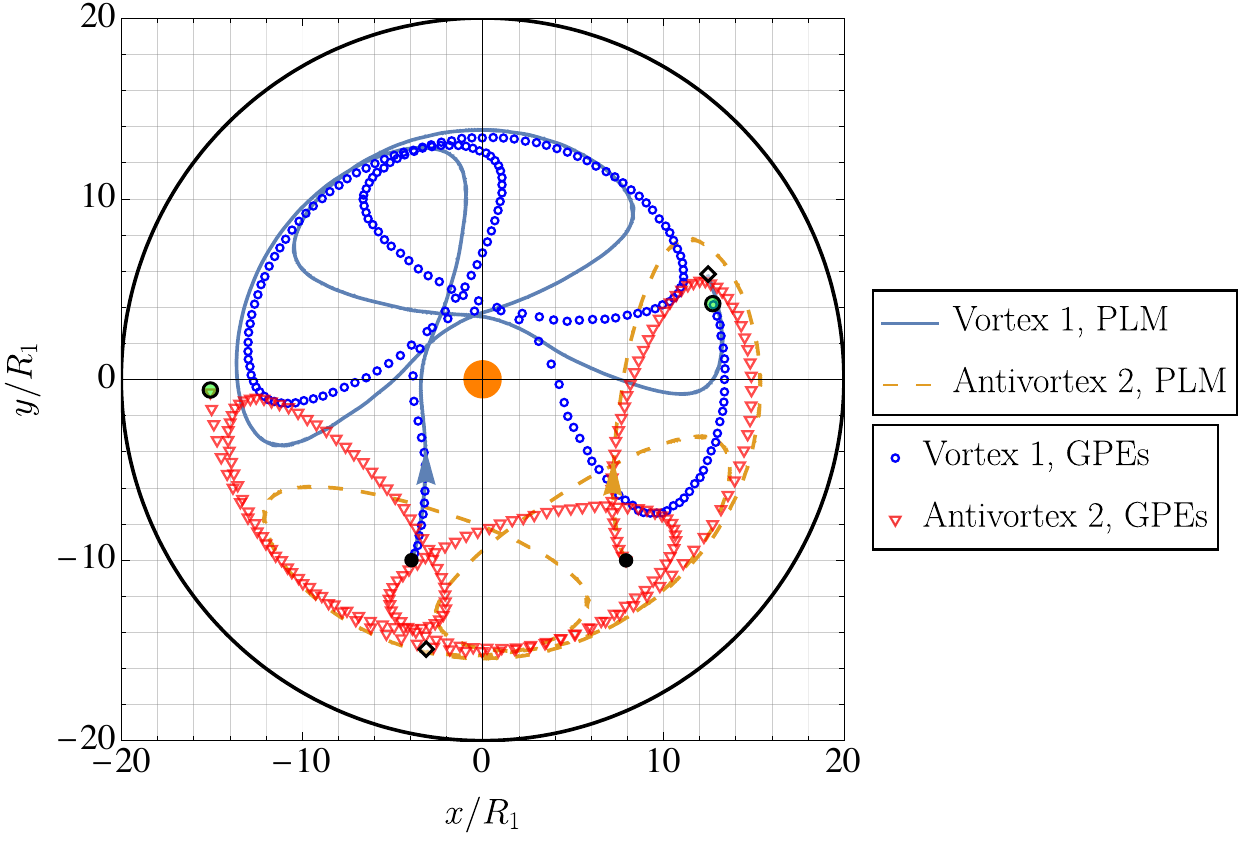}
\end{minipage}
\begin{minipage}{0.576\textwidth}
  \centering
  \includegraphics[width=1\linewidth]{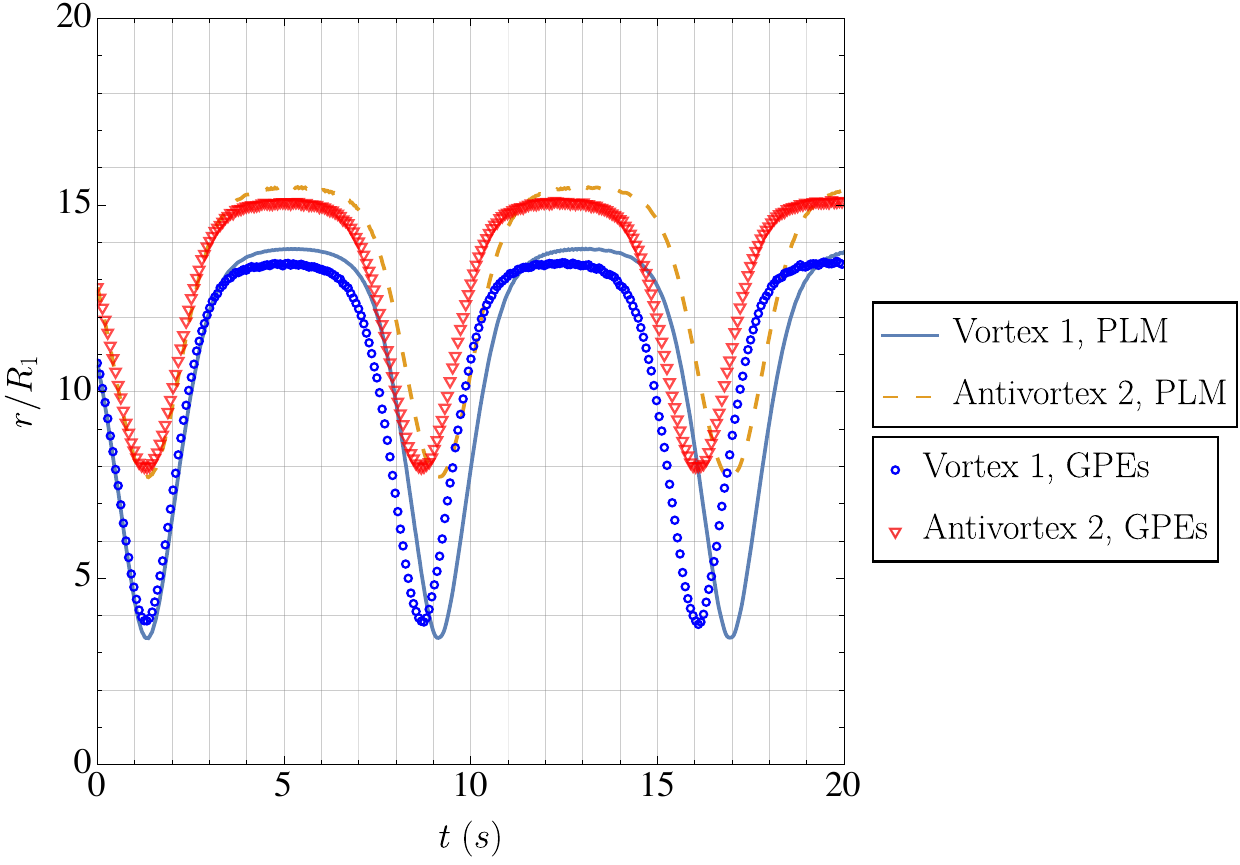}
\end{minipage}
\caption{Asymmetric scattering of the vortex dipole in a confined domain, in a case of go-around scenario. The GPEs results and the PLM prediction are qualitatively in agreement. The first two panels represent the vortex trajectories in the confining disk, whereas the last panel shows the plot of the radial positions evolution. We have that $N_b=5\times10^2$, and 
$\dot{x}_1(0)=\dot{x}_2(0)=0$, $\dot{y}_1(0)=\dot{y}_2(0)\simeq 2.4\times 10^{-5}$ $m/s$. 
The evolution time in the first (second and third) panel is of $8.5$ ($20$) $s$.}
\label{fig:asymm_go_around}  
\end{figure*}

{\it Asymmetric initial conditions: go-around scenario}. 
In the case of the go-around scenario illustrated in Figure \ref{fig:asymm_go_around}, it is impressive to see how well the PLM 
still captures the GPEs results, even at extremely long times. Qualitatively, the dynamical regime is fully captured, while we see again that the quantitative discrepancy arises as an effect of the boundary. As in the previous cases, the GP trajectories
stay globally further from the boundaries in comparison to the PLM prediction, as they enter the turning process induced by the boundary earlier than the PLM counterpart. Again, we can distinguish an increasing delay in the PLM positions with respect to the GPEs, best visible when comparing the radial coordinates of the vortices (last panel of Fig. \ref{fig:asymm_go_around}). 
Here it is evident how an approximately periodic character emerges in the scattering dynamics, present both in the ``real'' GP results and in the PLM's.
\\Note that, due to the asymmetry in the two vortex trajectories, also the delay of the PLM with respect to the GPEs results is different for the two vortices. This is well portrayed by the green circles and the white diamonds in the first two panels of Fig. \ref{fig:asymm_go_around}, representing the vortex final positions respectively according to the
GPEs and to the PLM.

\begin{figure*}[ht]
  \centering
  \begin{minipage}{0.5\textwidth}
    \centering
    \includegraphics[width=0.7\linewidth]{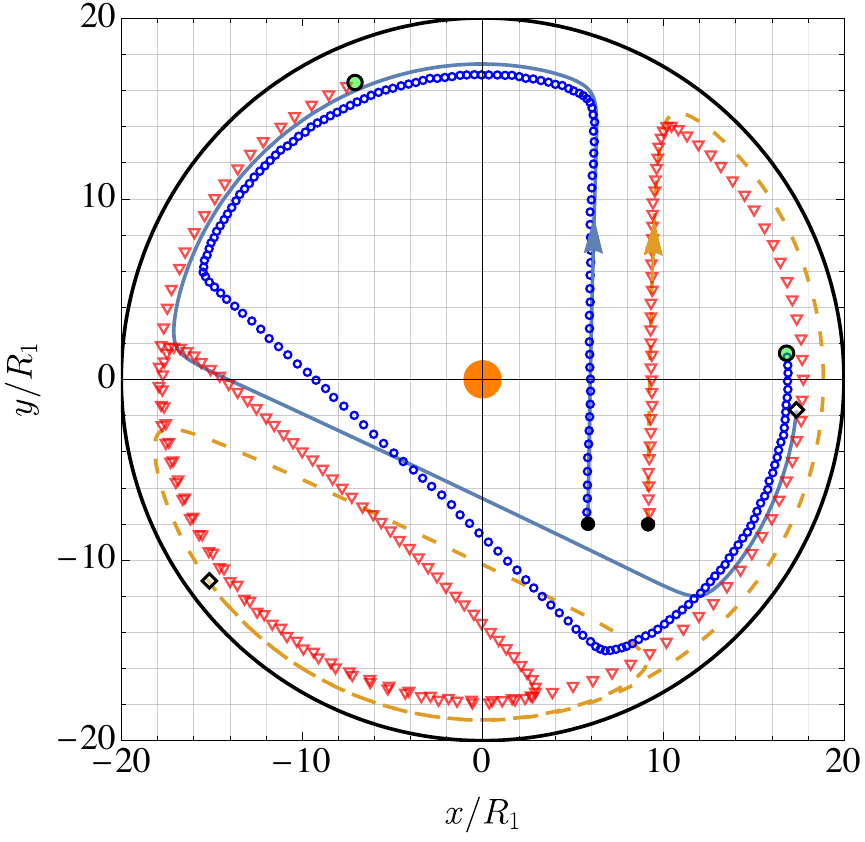}

    \vspace{1em}
    \includegraphics[width=0.7\linewidth]{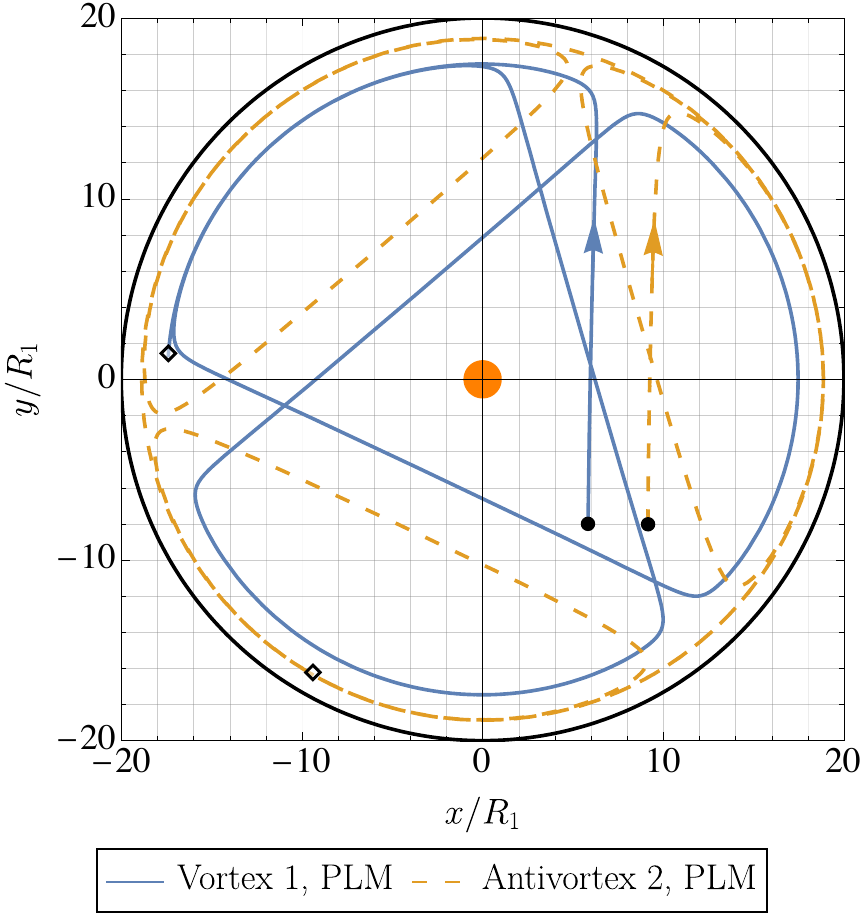}
  \end{minipage}%
  \hfill
  \begin{minipage}{0.5\textwidth}
    \centering
    \includegraphics[width=\linewidth]{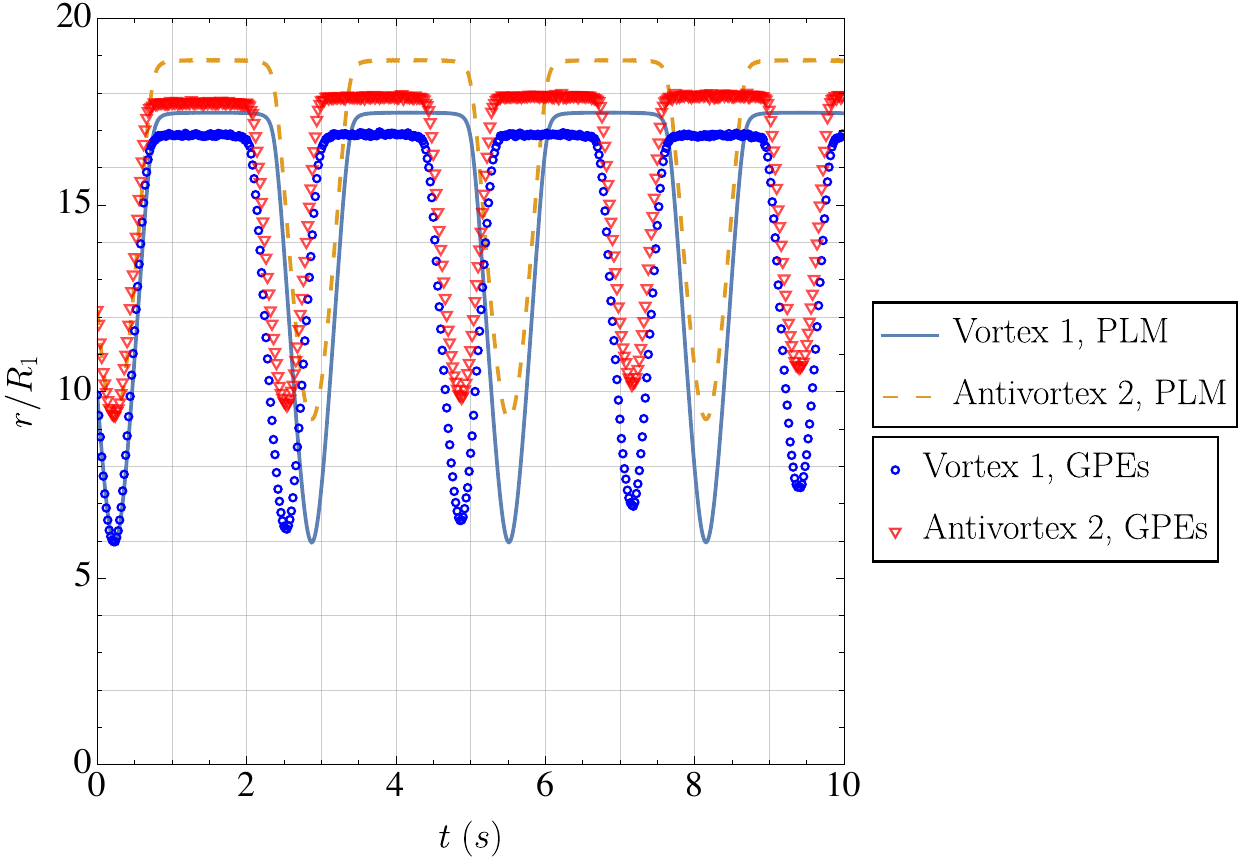}

    \vspace{1em}
    \raggedright
    \includegraphics[width=0.7\linewidth]{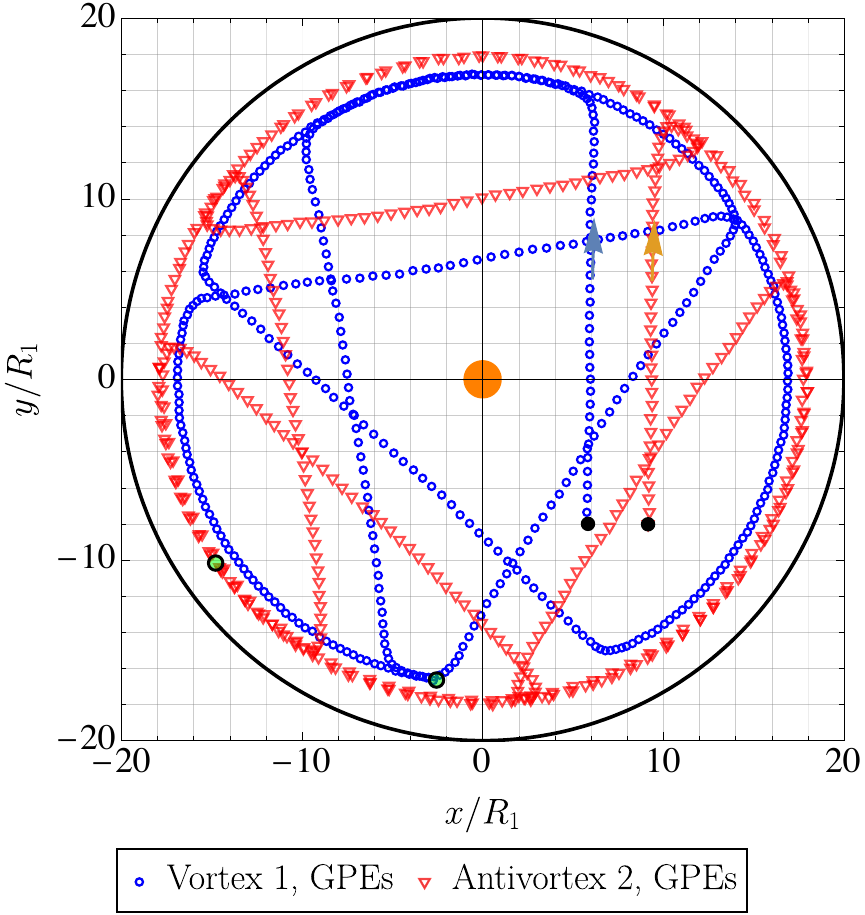}
  \end{minipage} 

  \caption{Fly-by instance of massive-dipole scattering in a confined domain. In the first panel we compare the GP trajectories with those of the PLM over an evolution time of $4$ $s$, whereas in the second panel we compare the vortices radial positions; the last two panels show the longer-time ($10$ $s$) trajectories according to the PLM and the GPEs. We have that 
  $\dot{x}_1(0)=\dot{x}_2(0)=0$, and $\dot{y}_1(0)=\dot{y}_2(0)\simeq 6.4\times 10^{-5}$ $m/s$.
  }
  \label{fig:fly_by_neat}
\end{figure*}

{\it Asymmetric initial conditions: fly-by scenario}.
As portrayed in Figure \ref{fig:fly_by_neat}, the fly-by type of scattering is also present in the GPEs solutions. Consistently with the previous cases, the black dots represent the vortices initial positions, while the white diamonds and green circles are relevant to the vortices final positions respectively according to the PLM and the GP results.
We see that the GP trajectories qualitatively reproduce the PLM's prediction. The third and fourth panels of the figure show respectively the PLM and the GP trajectory of the dipole over $10$ $s$ of time. 
In this case, unlike the go-around scenario,
two types of quantitative discrepancies arise.

The first type arises again as a consequence of an earlier interaction of the vortices with the boundary, and in case of Fig. \ref{fig:fly_by_neat}
translates into earlier recombination events in the GP dynamics with respect to the PLM.
The boundary effect is well visible in the 
shorter-time evolution in the
first panel of Fig. \ref{fig:fly_by_neat}, where it is evident how the dipole splitting due to the external boundary occurs earlier in the GP trajectory. The second panel of Fig. \ref{fig:fly_by_neat} shows the radial positions of the vortices over a longer time ($10$ $s$). In this panel, the plateaus are indicative of a circular trajectory in proximity of the boundary, as the two vortices are separated, whereas the wells represent the translating motion of the dipole after the recombination events. With respect to the PLM, the GP trajectories feature more translation segments. 

The second type of quantitative discrepancy between the GPEs and the model is an amplitude modulation in the otherwise essentially periodic oscillations of the radial vortex-positions (see the second panel of Fig. \ref{fig:fly_by_neat}). This arises in spite of the constant character of the dipole size during the translation segments, as well as the constant character of the radial position of the vortices during the circular motion arcs.
As a result, the GP solution features translating dipoles that grow progressively farther from the center over the simulated time, unlike the PLM prediction.
Interestingly, we observe a correlation between the damping of the oscillations in the vortex radial coordinates and the oscillations of the vortex-mass imbalance as the dipole translates. 
This vortex-mass imbalance corresponds to
a Josephson-like effect 
between the vortex cores, due to the two vortices of Fig. \ref{fig:fly_by_neat} being close enough to each other \cite{Bellettini2024PRR}. 
Different translation intervals are related to different types of oscillations in the vortex population imbalance, as visible in Figure \ref{fig:flyBy_imbalance}. 
This phenomenon suggests an interesting coupling between the bosonic transfer dynamics and the massive vortex motion that deserves a deeper future investigation.

As discussed in Sec. \ref{sec:PLM_annulus}, we can say that the boundary induces a slowing-down of the farther vortex, similar to what discussed in Sec. \ref{sec:PLM_plane} relative to the effect of the obstruction (see Eqs. \ref{dph12app}).

\subsection{Beyond-point-like effects}

\begin{figure}[h]
\centering
\begin{minipage}{0.5\textwidth}
  \centering
  \includegraphics[width=\linewidth]{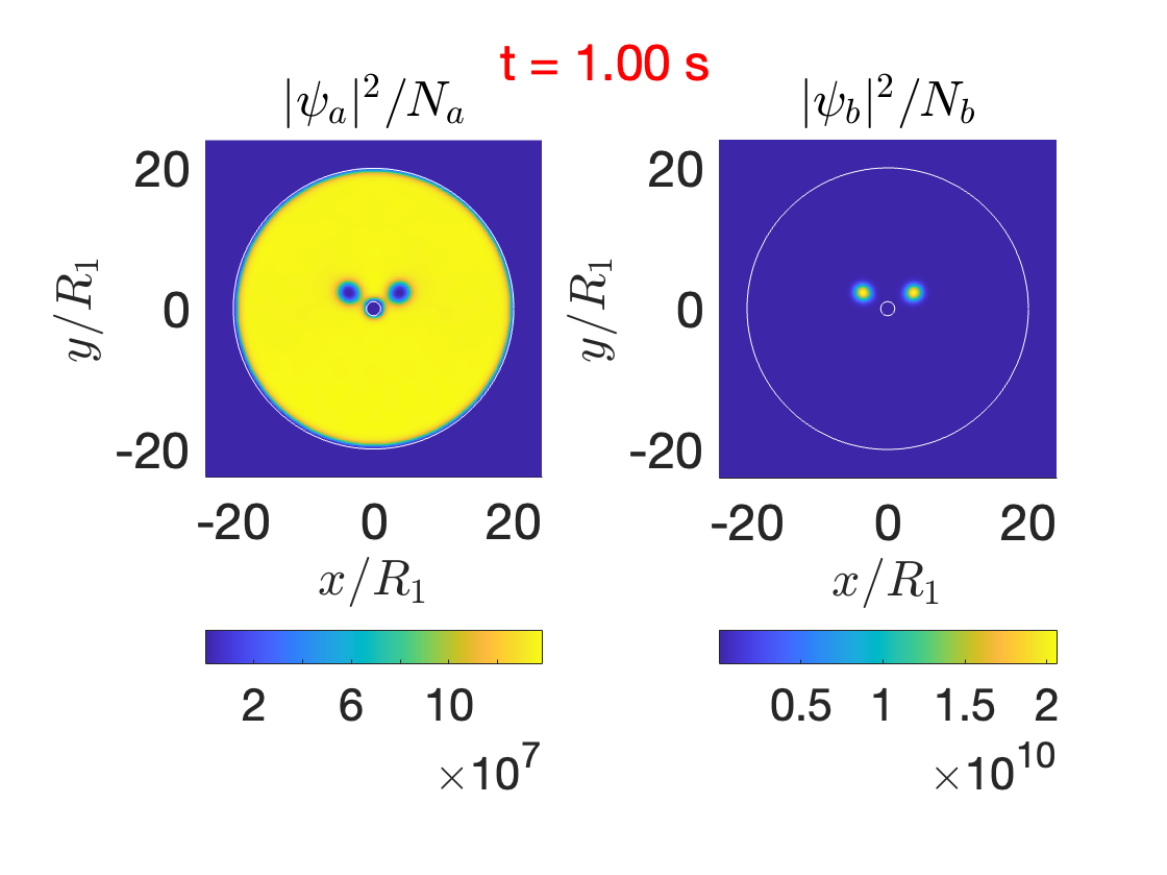}
\end{minipage}
\begin{minipage}{0.5\textwidth}
  \centering
  \includegraphics[width=\linewidth]{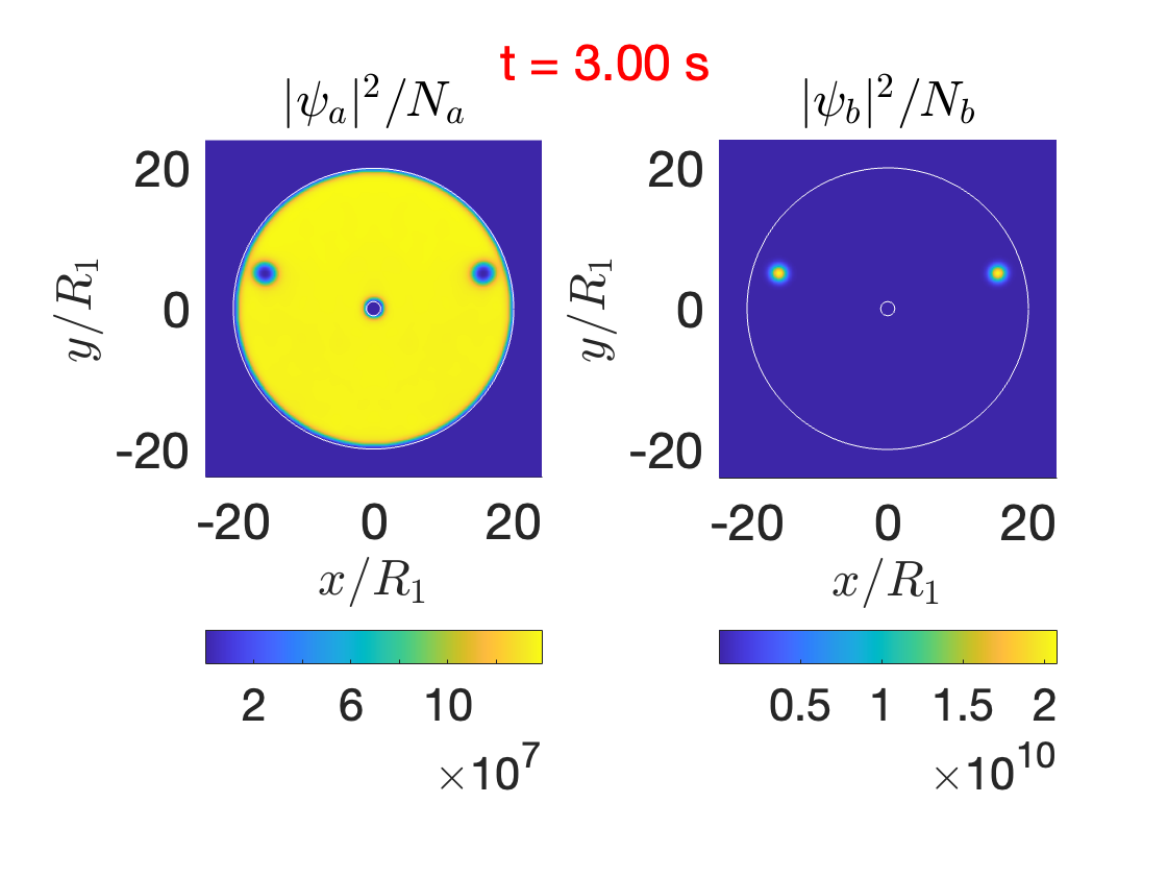}
\end{minipage}
\caption{Snapshots of the number density fields $|\psi_a|^2$ and $|\psi_b|^2$ during the time evolution of the GPEs, at different times, for a symmetric dipole scattering. The vortices trajectories are those of the right panel in Fig. \ref{fig:symm_trajs_comp}.}
\label{fig:symm_snapshots}  
\end{figure}

It is interesting to discuss the beyond-point-like phenomenology characterizing the GPEs solution, to explain the discrepancies in the GP and PLM results. In the previous section, we observed how the discrepancies between model and mean-field picture seem to arise in correspondence of boundaries, such as that of the obstruction or that of the confining disk. A possible explanation for the different sensitiveness of the GPEs solution versus the PLM to the boundaries consists in considering the role of the density profile $|\psi_a|^2$ at the vortex sites and close to the boundaries. This is approximated by
the constant $n_a$ almost everywhere (see Eq. \eqref{eq:psi_a}) at the level of the PLM. Conversely, the real GP profile involves a density well at a vortex core's site \cite{Fetter2009}, 
and a depletion at the boundaries leading to a smooth transition from the bulk value to zero. These features are visible in the snapshots in Figure
\ref{fig:symm_snapshots}, relevant to the same trajectory of the right panel of Fig. \ref{fig:symm_trajs_comp}.
The figure shows how the depletion at the vortex cores and at the boundaries may overlap,
so to induce a density profile that
is not even locally symmetric with respect to the vortex center.
Most likely due to such non-point-like character of the field solution, the latter is affected by a stronger sensitivity to the boundary. 
Note that the larger the vortex mass accommodated by the core, the larger the latter is, and the more
important its finite size effects are.

\begin{figure}[h]
\centering
\includegraphics[width=1\linewidth]{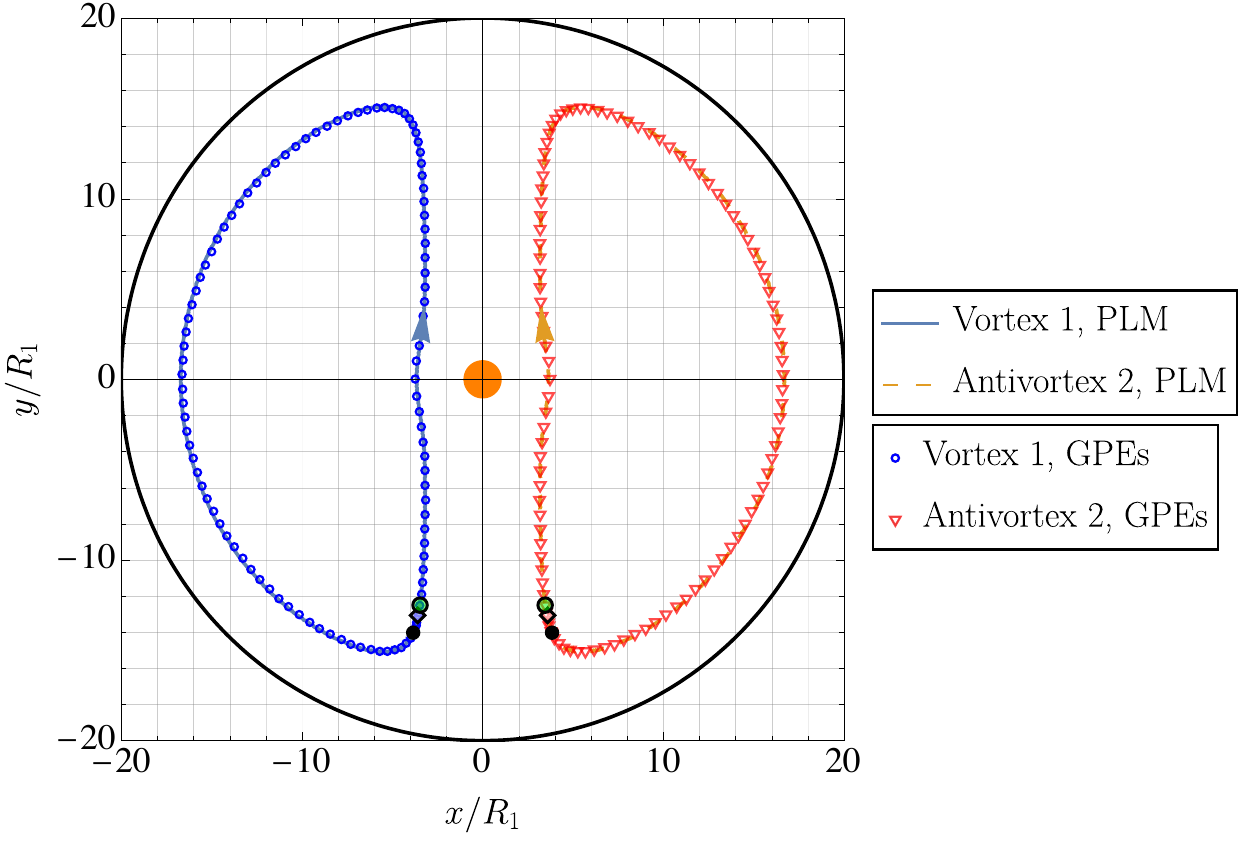}
\caption{Symmetric scattering of a vortex dipole against an obstacle in a confined domain, and comparison with the GPEs, for a system at values of $g_a$ and $g_{ab}$ that are $10$ times the standard ones (see Sec. \ref{sec:PL_model}). The other physical parameters are the same of the right panel of Fig. \ref{fig:symm_trajs_comp}, and the simulation time if of $5.4$ $s$.} 
\label{fig:symm_RbK_10}  
\end{figure}

\begin{figure}[h]
\centering
\begin{minipage}{0.5\textwidth}
  \centering
  \includegraphics[width=\linewidth]{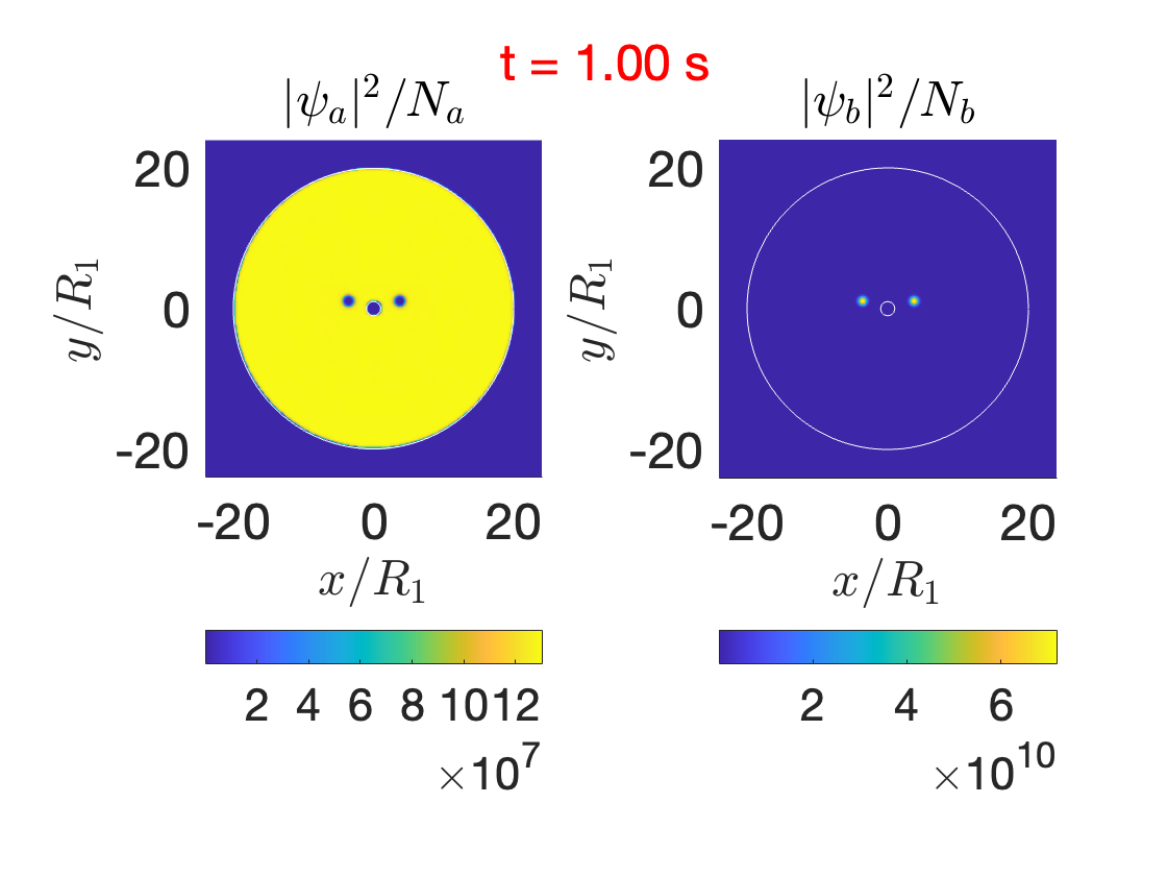}
\end{minipage}
\begin{minipage}{0.5\textwidth}
  \centering
  \includegraphics[width=\linewidth]{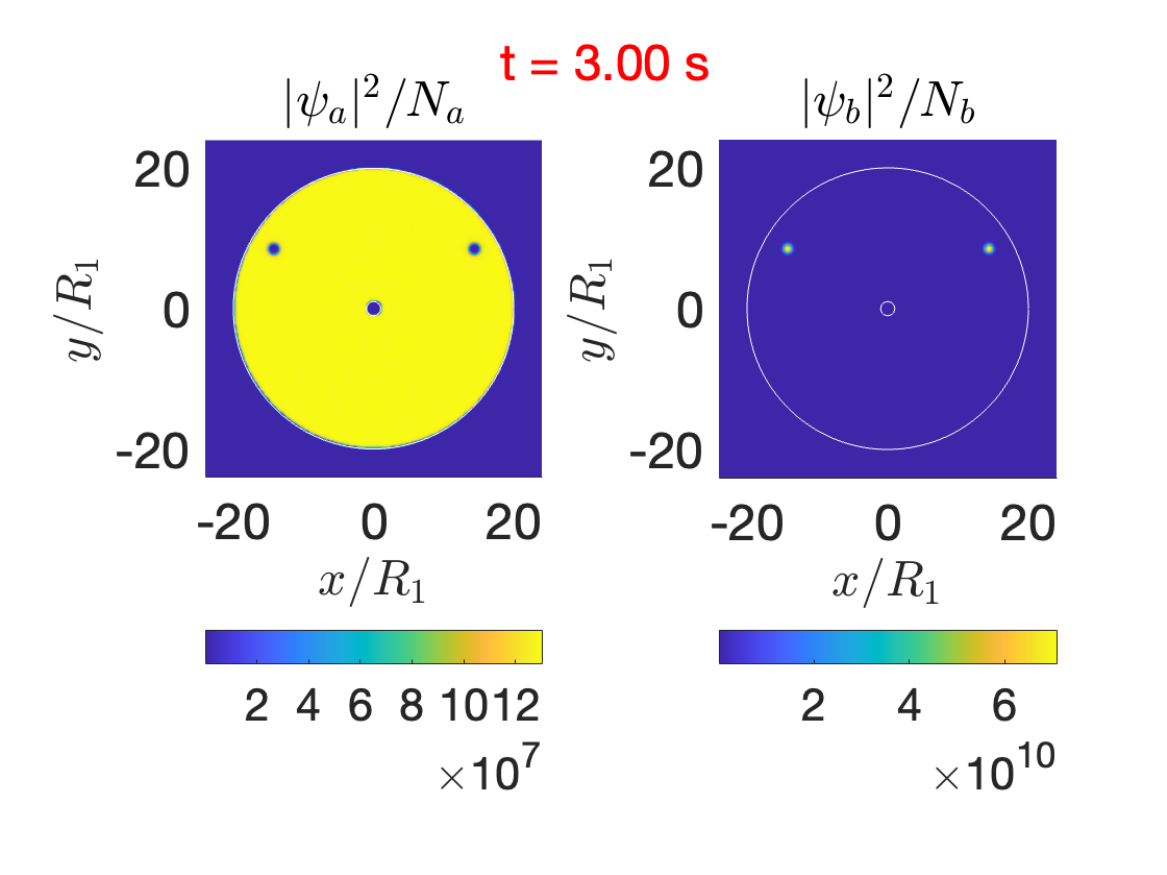}
\end{minipage}
\caption{Snapshots of $|\psi_a|^2$ and $|\psi_b|^2$ during the time evolution of the GPEs for the symmetric dipole scattering illustrated in Fig. \ref{fig:symm_RbK_10}.}
\label{fig:symm_snapshots_10}  
\end{figure}

This hypothesis is supported by the results obtained in a pedagogical example at much higher repulsion coefficients $g_a$ and $g_{ab}$ with respect to the standard values. In this case, both the vortex cores and the $b$-infilling peaks are much narrower, as visible in Figure \ref{fig:symm_snapshots_10}. 
As illustrated in Fig. \ref{fig:symm_RbK_10}, a direct consequence of such size effects is a perfect agreement
between GP and PLM solutions.
This result is to be compared with
the trajectory in the right panel of Fig. \ref{fig:symm_trajs_comp}, featuring lower repulsion coefficients and a weaker agreement between PLM and GPEs.
In the future, it would be interesting to study whether an improvement of the PLM \textit{Ans{\"a}tze} to take into account the non-constant density profile of $a$ (see e.g. Ref. \cite{Bellettini2023}) could lead to a more precise description of the near boundary dynamics.

Other beyond-point-like effects featured by the GPEs solutions include the tunneling of the $b$ component between the two vortex wells (see Refs. \cite{Bellettini2024PRR} and \cite{Pola2012}), and the emergence of sound waves within the fluid $a$ in relation to vortex accelerations \cite{Barenghi2005}. The $b$-tunneling, emerging at low vortex distances, is responsible for a time-dependent variation of the vortex masses, which affects the vortex dynamics e.g. by inducing accelerations or an expansion and compression of the vortex core. 
On the other hand, also the trajectories that are not captured by the PLM, as they involve divergencies in the equations of motion, correspond to beyond-point-like effects in the GPEs. These consist in general in the disruption of the massive vortices in the sense of the PLM (see the \textit{Ans{\"a}tze} \eqref{eq:psi_a} and \eqref{eq:psi_b}). In this case, phenomena as two-vortex overlaps or a vortex-boundary overlap may occur, as well as
the absorption of a vortex at the obstacle, and the conversion of the former into a quantized background supercurrent, with the expulsion of the vortex mass from the core.

\section{Conclusions}
\label{sec:concl}

In the present paper, we investigate the scattering mechanisms that 2D massive quantum vortex dipoles undergo in presence of an obstacle.

We start with the characterization of the scattering mechanisms in an infinite plane geometry, i.e. where the effects of any confining traps are negligible. 
Our first result, analogous to what found numerically
in Ref. \cite{Griffin2017} for massless vortices, is the identification of two fundamental scattering behaviors by means of a point-like model, introduced in Sec. \ref{sec:PL_model}. The predictions of such analytical model are validated by the numerical simulation of coupled Gross-Pitaevskii equations. 

These fundamental behaviors, treated in Sec. \ref{sec:PLM_plane}, have in common the effective deviation of the dipole translation line, and are referred to as the go-around and the fly-by. The former is characterized by a temporary capturing of the closest vortex to the obstacle by the obstacle itself, with the destruction of the dipole and the interruption of its translation. Subsequently, the two vortices recombine in a dipole, and enter a new translating trajectory. This final translation is in general deviated with respect to the dipole translation that precedes the scattering event.
On the other hand, trajectories in the fly-by behavior are marked by a deviation, due to the obstacle, of the dipole translation line, with no temporary destruction of the dipole configuration.
Whether a dipole enters the first or second behavior is discriminated by 
the combination of pair interactions between real and virtual vortices caused by the boundary. Interestingly, both the scattering behaviors highlight the stability of the dipole vortex-state.

Another interesting result, in Sec. \ref{sec:free_dipole},
concerns the effect of the vortex mass on the trajectories of the dipoles.
We find that, far enough from the obstacle, a perturbative treatment can be applied to the VA-pair equations which, at the zero order, highlights a purely dipole-like behavior of the VA pair. This corresponds to a subregime characterized by vortex trajectories whose distinctive feature is to exhibit the same initial velocities. While this subregime appears to be independent from the vortex masses, the role of the latter is recovered by the first-order dynamics, which describes the small vortex-mass oscillations
(see, for example, Fig. \ref{fig:PL_asymmetrical_so}).
Basing on the numerical solutions of PLM equations, 
we then confirm that, in the massive-dipole regime, {\it i}) far enough from the obstacle the massive-dipole dynamics collapses to a
massless-dipole dynamics,
and {\it ii}) also during the interaction with the obstacle, the vortex trajectories seem to be indistinguishable from those of massless vortices. Nevertheless, the effect of vortex masses is then recovered and analyzed in Fig. \ref{fig:massless} showing how they can influence the dipole deviations in the scattering process.

In Section \ref{sec:scatt}, we also find some analytic formulas for the discrimination of the two scattering behaviors, given in terms of the minimal distance from the obstacles that is reached by the two vortices
during their dynamical evolution.

Our analysis of the scattering processes involving a vortex dipole in presence an obstacle, suggests the very intriguing idea of controlling the scattering angle.
This potential application of the PLM is summed up for example in
the
plot of the deflection angle of Fig. \ref{fig:theta_h_100}.

In Section \ref{sec:PLM_annulus} we treat by means of a PLM the scattering phenomenology of a dipole in a confined domain, where both the external circular trap and the obstacle are centered in the origin.
We observe similar scattering behaviors as in the case of the plane-like domain, where however significant quantitative differences arise due to the effect of the external boundary. Furthermore, in the proximity of the external boundary, there is no dipole translation, upon the contrary the dipole is destroyed, and the two vortices follow an almost circular motion
along the disk boundary until they recombine in a translating dipole.
Likewise to what happens in correspondence of the obstacle, during the circular orbits of the two vortices alongside the external boundary, we have that the vortex closer to the boundary moves faster than the other.
Our results highlight a cyclic dynamics of the VA pair within the confined domain, characterized by a repeated sequence of a translation segment, a scattering event followed by the eventual dipole recombination, and finally the dipole splitting at the external boundary, to then reach a final dipole recombination before starting a new sequence. Remarkably, in this processes, the dipole size (as it translates) is always the same. 

Further on, in Section \ref{sec:GPE}, we investigate the scattering events according to the mean-field dynamics determined by the Gross-Pitaevskii equations. We take the more complex (and realistic) case of a trapped BEC mixture and compare the results with the predictions of the PLM. In spite of the considerably more complex geometry characterizing the annulus, our results show a very good qualitative
agreement between PLM and GPEs, where the discrepancies arise close to the boundaries likely as a consequence of the
non-constant character of the density fields. 
To corroborate this assumption we test the PLM against the GPEs results at larger $g_a$ and $g_{ab}$, so to guarantee narrower vortex cores in the fields simulation. Remarkably, we observe a substantial improvement in the agreement between PLM and GPEs.

As the PLM Lagrangian of the annular geometry tends to that of the plane in the limit of large $R_2$, a similar or better agreement between GPEs and PLM is expected for the plane.

As a conclusion, even in the absence of analytical solutions describing the dipole scattering trajectories, thanks to the PLM we could discriminate two classes of trajectories, or scattering processes and some of their features. 
The theoretical approach developed in this paper is directly applicable for the case of many dipoles, of interest as a future outlook.
Moreover, a direct extension of the present framework is the case of an unbalanced dipole, i.e. of
different vortex masses.
In this context, another future perspective is the inclusion, in the PLM, of varying vortex masses, to try to capture the tunneling phenomena featured by the GPEs solutions at close vortex distances \cite{Meng2025}. 
Finally, it would be interesting to extend our analysis to investigate
the effect of massive cores in the dynamics of vortices in Fermi superfluids and in the
dissipative effects thereof \cite{Grani2025}.

\newpage

\begin{appendix}

\section{Dipole equations}
\label{A1}

The polar-coordinate picture of the equations of motion
\eqref{eq:eq_rho}
and
\eqref{eq:eq_theta} for a VA pair reads   

$$
m_1 \Bigl ( {\ddot {r}}_1 - r_1 {\dot \theta}_1^2 \Bigr )
+
\rho_* k r_1 {\dot \theta}_1
= \frac{\rho_* k^2}{2\pi} \times
$$
\begin{equation}
\left [ -\frac{ r_1-r_2 c_{12}}{r_{12}^2 } 
+\frac{r_2^2r_1 -r_2 c_{12} R_1^2}{D} -\frac{r_1}{r_1^2 -R_1^2} 
\right ],
\label{eqr1}
\end{equation}
\begin{equation}
\begin{split}
m_1 \Bigl ( 2{\dot {r}}_1 {\dot \theta}_1 + r_1 {\ddot \theta}_1 \Bigr )
-&
\rho_* k {\dot r}_1
\\
&=
\frac{\rho_* k^2}{2\pi} \left ( \frac{R_1^2}{D} - \frac{1}{r^2_{12}}  
\right ) r_2 s_{12},
\end{split}
\label{eqth1}
\end{equation}

$$
m_2 \Bigl ( {\ddot {r}}_2 - r_2 {\dot \theta}_2^2 \Bigr )
-
\rho_* k r_2 {\dot \theta}_2
=
$$
\begin{equation}
\frac{\rho_* k^2}{2\pi} \left [ -\frac{ r_2-r_1 c_{21}}{r_{12}^2 } 
+\frac{r_1^2r_2 - r_1 c_{21} }{D} -\frac{r_2}{r_2^2 -R_1^2} 
\right ],
\label{eqr2}
\end{equation}
\begin{equation}
\begin{split}
m_2 \Bigl ( 2{\dot {r}}_2 {\dot \theta}_2 + r_2 {\ddot \theta}_2 \Bigr )
+&
\rho_* k {\dot r}_2
\\
&=
\frac{\rho_* k^2}{2\pi} \left ( \frac{R_1^2}{D} - \frac{1}{r^2_{12}} 
\right ) r_1 s_{21},
\end{split}
\label{eqth2}
\end{equation}
with $c_{12} = \cos (\theta_1-\theta_2)= c_{21}$, $s_{12} = \sin (\theta_1-\theta_2)= -s_{21}$, $D= r_1^2 r_2^2 +R_1^4- 2R_1^2 {\vec r}_1 \cdot {\vec r}_2$ and $r_{12}= |{\vec r}_1- {\vec r}_2 |$. 

The radial components of the previous equations provide equation \eqref{radeq12c1} whose solution allows us to identify the angular velocities \eqref{dphi1} and \eqref{dphi2}.
Interestingly, for $d_1 \simeq R_1$ and a generic $d_2 > R_1$ one easily proves that
$$
\frac{ 1}{r_1} A(r_1,r_2) \simeq \frac{r_1}{r_1^2 -R_1^2}, 
\quad 
\frac{ 1}{r_2} A(r_2,r_1) \simeq \frac{r_2}{r_2^2 -R_1^2},
$$
implying, in turn,
\begin{equation}
\omega^-_1 
\simeq
\frac{\rho_* k }{ 2m_1 } 
\left [ 1 -
\sqrt{ 1 + \frac{2m_1}{\pi \rho_* ( d_1^2 -R_1^2) } }
\right ] < 0 ,
\label{dph1A}
\end{equation}
\begin{equation}
\omega^+_2 
\simeq
\frac{\rho_* k }{ 2m_2 } 
\left [ -1 + \sqrt{ 1+ \frac{2m_2}{\pi \rho_* (d_2^2 -R_1^2)} }
\right ]\; >  0,
\label{dph2A}
\end{equation}
In the case 2, we easily derive from equations \eqref{Omega1} and \eqref{Omega2} the approximated formulas for the angular velocities
in the limiting case $d_i, d_2 >> R_1$ 
\begin{equation}
{\Omega}_1  
\simeq
\frac{\rho_* k}{2m_1} \left [ 1
- \sqrt{ 1 - \frac{2m_1}{\pi \rho_*} \frac{ 1}{d_1(d_2-d_1)} 
}
\right ] 
\label{dph1B}
\end{equation}

\begin{equation}
{\Omega}_2  
\simeq
\frac{\rho_* k}{2m_2} \left [ -1
+ \sqrt{ 1 + \frac{ 2m_2}{\pi \rho_*} \;\frac{ 1}{r_2 |r_1-r_2|}
}
\right ].
\label{dph2B}
\end{equation}
The definition of the effective masses
$$
M_1' = \pi \rho_* (d^2_2- d^2_1) \frac{d_1}{d_2+d_1}, \,
M_2'= \pi \rho_* (d^2_2- d^2_1) \frac{d_2}{d_2+d_1},
$$
implementing the conditions $M_1' >> m_1$ and $M_2' >> m_2$ leads, in turn, to the elementary formulas \eqref{Om12}. Note that these conditions are easily realized when considering the strong imbalance of vortex masses with respect to the mass of the annulus with area $\pi \rho_* (d^2_2- d^2_1)$.
\medskip

\noindent
{\it Transition to the regime} $\Omega^-_1 <0$. Here, we discuss in a more detailed way the dependence of $\Omega_1^-$ from its parameters. Formula \eqref{Omega1} (with the minus sign) provides a positive angular velocity under the double condition
\begin{equation}
0 < 1 - \frac{2m_1}{\pi \rho_* d_1} B(d_1,d_2)  < 1.
\label{2ineq}
\end{equation}
The right inequality reduces to $B(d_1,d_2) >0$, the crucial condition ensuring that $\Omega_1^- >0$ is physically acceptable. Nevertheless, the left equality is also important because excludes complex-valued angular velocities. The left inequality becomes
$$
\frac{2m_1}{\pi \rho_* d_1} \left ( \frac{ 1}{d_2-d_1} +\frac{d_2}{d_1d_2 -R_1^2} +\frac{d_1}{d_1^2 -R_1^2} \right ) < 1.
$$
This could be violated either for $d_1$ sufficiently close to $R_1$ or for $d_1$ sufficiently close to $d_2$. In the first case, the inequality is satisfied if $M_R =\pi \rho_* R_1^2 > m_1$ is satisfied. In the second case an upper critical value $R_1'$ for $d_1$ can be found (numerically) which guarantees that the inequality is satisfied for $d_1 < R_1'$. If $d_2 >> R_1$  the approximate formula for this value is given by 
\begin{equation}
R_1' \simeq d_2 \sqrt{ 1 -4
\frac{m_1}{M_2} },
\label{Runo}
\end{equation}
showing how for $M_2 =\pi \rho_* d_2^2 >> m_1$ (a condition that can be easily realized) one has $R_1' \simeq d_2$.

\begin{figure}
    \centering
    \includegraphics[width=\linewidth]{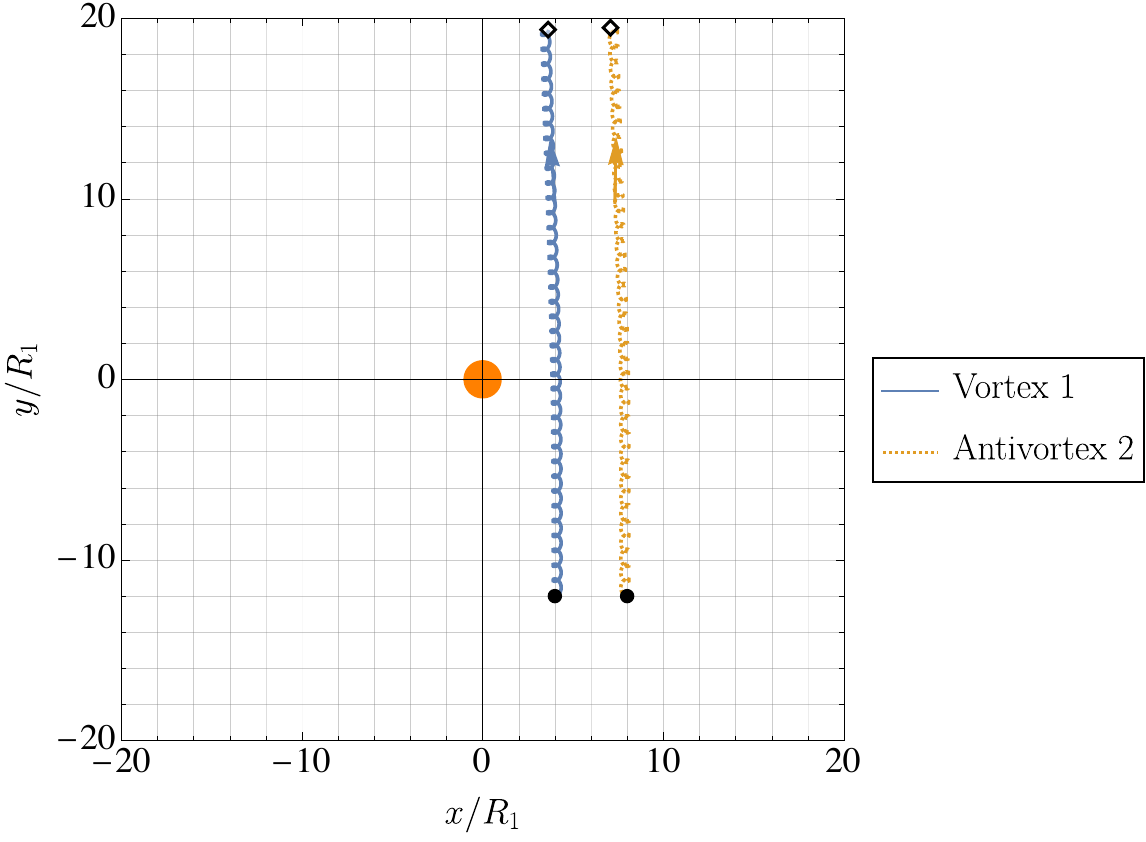}
    \caption{Dipole trajectory within an unbounded domain, with a scattering event, featuring visible small oscillations caused by the vortex mass. 
    In this case, $N_b=  10^3$, the initial velocities are $\dot{x}_1(0)=-\dot{x}_2(0)\simeq 1.1\times 10^{-4}$, $\dot{y}_1(0)=\dot{y}_2(0)=0$ $m/s$,
    and the evolution time is $t=1$ $s$.}
    \label{fig:PL_asymmetrical_so}
\end{figure}


\label{Algebra}

\section{Additional figures}

\begin{figure*}
\centering

\begin{minipage}{0.32\textwidth}
  \centering
  \includegraphics[width=\linewidth]{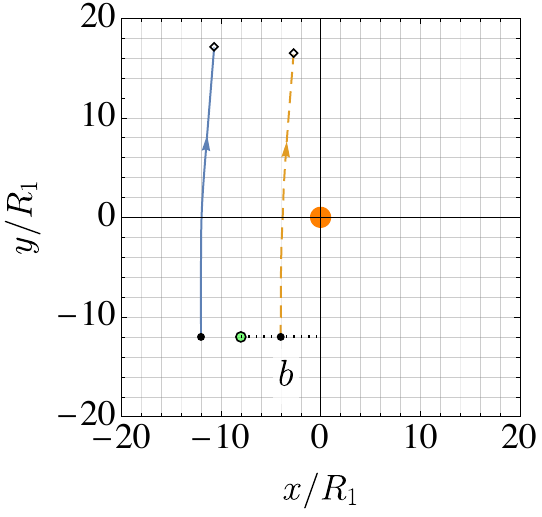}
\end{minipage}\hfill
\begin{minipage}{0.35\textwidth}
  \centering
  \raisebox{4.4em}{\includegraphics[width=\linewidth]{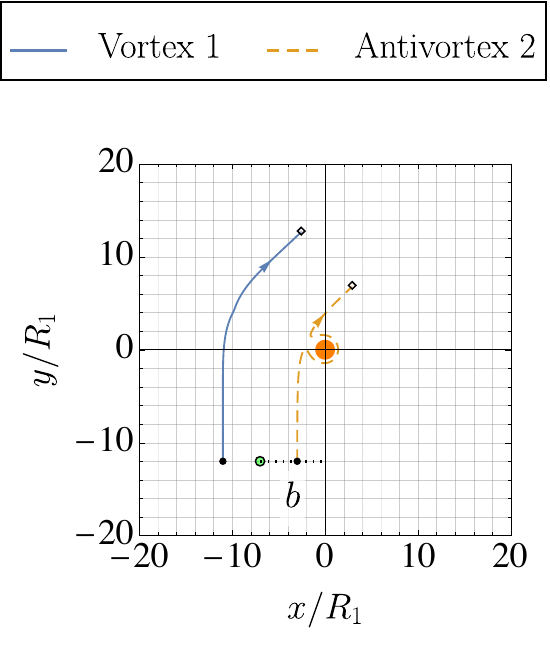}}
\end{minipage}\hfill
\begin{minipage}{0.32\textwidth}
  \centering
  \includegraphics[width=\linewidth]{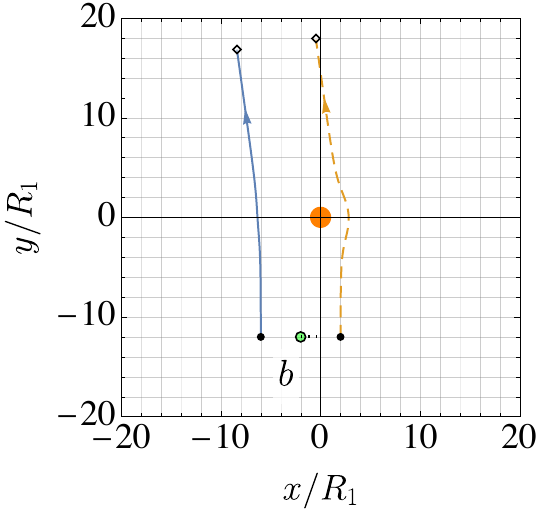}
\end{minipage}

\vspace{0.5em}

\begin{minipage}{0.33\textwidth}
  \centering
  \includegraphics[width=\linewidth]{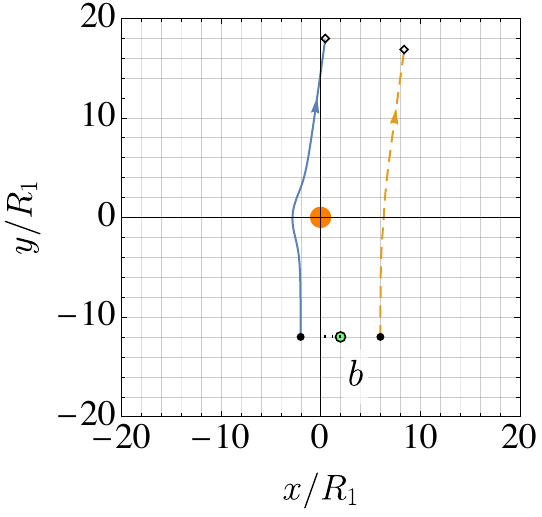}
\end{minipage}\hfill
\begin{minipage}{0.33\textwidth}
  \centering
  \includegraphics[width=\linewidth]{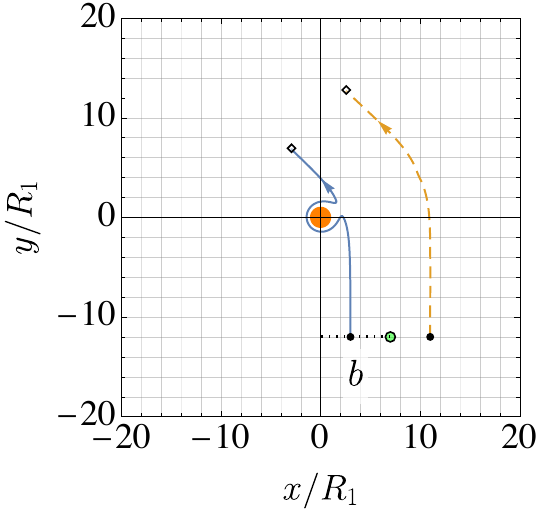}
\end{minipage}\hfill
\begin{minipage}{0.33\textwidth}
  \centering
  \includegraphics[width=\linewidth]{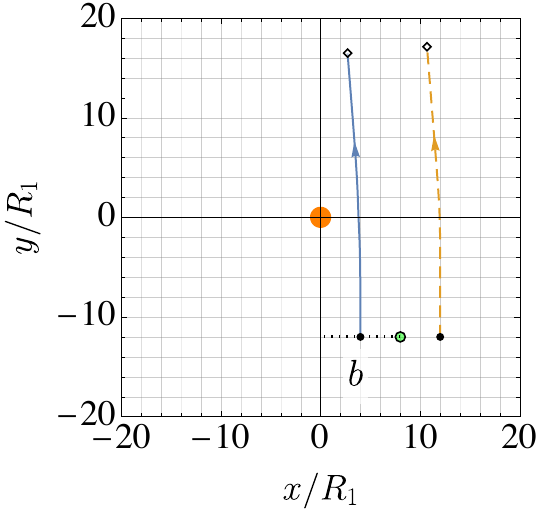}
\end{minipage}

\caption{Sequence of scattering events, for a vortex dipole in an unbounded domain, as the value of the impact parameter $b$ is progressively increased. The white-shaded squares mark the ending points of the trajectories. The initial velocities are $\dot{x}_1(0)=\dot{x}_2(0)=0$, and $\dot{y}_1(0)=\dot{y}_2(0)\simeq 3.7\times 10^{-5}$ $m/s$, and the evolution time is $t=2$ $s$.}
\label{fig:scattering_sequence}
\end{figure*}

In Figure \ref{fig:scattering_sequence} we propose a sequence of dipole trajectories to better illustrate the meaning of Fig. \ref{fig:theta_h_100}. In the panels, different types of scattering events are visible, originated at increasing values of the impact parameter $b$ (sketched through the $x$-coordinate of the green-shaded circle), while all the other physical quantities are kept constant.
In particular, it is shown how the dipole transits through a fly-by behavior (at large negative $b$), to then enter some go-around trajectory characterized by a large deflection angle, that goes decreasing in modulus at larger $b$. The third panel shows indeed a go-around scattering process characterized by a lower deflection angle $\phi$, in modulus, that is however positive in sign unlike in the preceding panels. Increasing $b$ up to zero, one retrieves a symmetric scattering of the type of Fig. \ref{fig:PL_symmetrical}. The panels of the second row show the symmetric, with respect to the origin, counterparts of the first row.

\begin{figure}
    \centering
    \includegraphics[width=\linewidth]{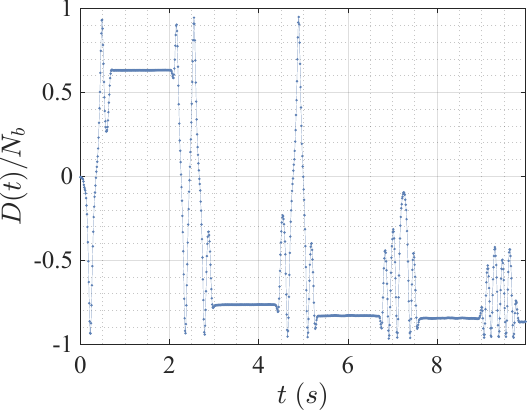}
    \caption{Variation of the population imbalance $D$ over time between the two vortices relevant to the evolution of Fig. \ref{fig:fly_by_neat}.}
    \label{fig:flyBy_imbalance}
\end{figure}

Figure \ref{fig:flyBy_imbalance} illustrates the population imbalance $D=(m_1-m_2)/m_b$ between the two vortex cores as a function of time, relevant to the system of Fig. \ref{fig:fly_by_neat}. By comparing the figure with the second panel of Fig. \ref{fig:fly_by_neat} one infers how the tunneling dynamics of the vortex infilling component is coupled with the VA dynamics. During the dipole translation segments, some oscillations of $D$ take place, indicating a mass exchange between the two vortex cores. On the other hand, as the VA move of circular motion along the boundary, the population imbalance $D$ is frozen at a given value, varying over time. 
The trend of $D$ as in Fig. \ref{fig:flyBy_imbalance} at long times suggests a saturation of the frozen value of $D$ when the dipole splits at the boundary.
As mentioned in Sec. \ref{sec:GPE_PLM}, Fig. \ref{fig:flyBy_imbalance} hints at an interesting correlation between the variation of $D$-oscillations and the damping of the oscillations in the radial coordinates of the vortices (see second panel of Fig. \ref{fig:fly_by_neat}). Namely, the $D$-oscillations pass from being very large in the first translation segments (Josephson-like) to being smaller in amplitude as in a self-trapped-like state of the $b$-subsystem.

\end{appendix}

\newpage


%

\end{document}